\documentclass[aps,a4paper, preprint, superscriptaddress,preprintnumbers,floatfix,nofootinbib,amsmath,amssymb]{revtex4}
\usepackage{url}
\usepackage{hyperref}
\usepackage{cleveref}

\usepackage{amstext,amssymb}
\usepackage{amsmath}
\usepackage{graphicx}
\usepackage{xspace}
\usepackage{color}
\usepackage{units}
\usepackage{slashed} 
\usepackage{multirow}
\newcommand{\be}{\begin{equation}}
\newcommand{\ee}{\end{equation}}
\newcommand{\bea}{\begin{eqnarray}}
\newcommand{\eea}{\end{eqnarray}}
\newcommand{\nn}{\nonumber}

\def\s1{\hat s}

\usepackage{slashed}

\newcommand{\nua}[1]{\ensuremath{\rlap{\kern-2.5pt\ensuremath{\overset{\scriptscriptstyle(-)}{\phantom{\nu}}}}{\ensuremath{{\nu}_{#1}}}}\xspace}

\begin{document}
\title{ Neutrino mixing and Leptogenesis with modular $S_3$ symmetry in the framework of type III seesaw }
\author{Subhasmita Mishra}
\email{subhasmita.mishra92@gmail.com}
\affiliation{Department of Physics, IIT Hyderabad, Kandi - 502285, India}
\begin{abstract}
Discrete symmetries being preferred to explain the neutrino phenomenology, we chose the simplest $S_3$ group and explore the implication of its modular form on neutrino masses and mixing. Non-trivial transformations of Yukawa couplings under this symmetry, make the model phenomenologically interesting by reducing the requirement of multiple scalar fields. This symmetry imposes a specific flavor structure to the neutrino mass matrix within the framework of less frequented type III seesaw mechanism and helps to explore the neutrino mixing consistent with the current observation. Apart, we also explain the preferred scenario of leptogenesis to explain the baryon asymmetry of the universe by generating the lepton asymmetry from the decay of heavy fermion triplet at TeV scale.
\end{abstract}

\maketitle
\flushbottom

\section{Introduction}
The success of standard model (SM) is limited to accommodate certain experimental observations like neutrino masses, matter-anti matter asymmetry and existence of dark sector etc \cite{Gripaios:2015gxa,Araki:2004mb,Tanabashi:2018oca,Ade:2015xua}. Therefore the extension of the SM particle spectrum is necessary to explain those limitations. Discrete symmetries are proven to be more fruitful in this direction, since imposition of these symmetries provides a specific flavor structure to the neutrino mass matrix and hence being widely used in neutrino phenomenology \cite{King:2015bja,Chattopadhyay:2017zvs,Sartori:1979gt,Wilczek:1977uh,DeRujula:1977dmn,Haba:2005ds,Altarelli:2010gt}. Few examples are $S_3$, $A_4$ and $S_4$ etc., which are found to be commonly used in the literature \cite{Petcov:2018snn,Kubo:2004ps,Borah:2017dmk,Ishimori:2010au}. But these discrete groups always require the inclusion of multiple scalar fields with specific alignment of vacuum expectation values (VEV). Such complications can be avoided by the nontrivial transformation of Yukawa couplings under these symmetries and this idea has been achieved a decent attention and well explored in literature as modular symmetries \cite{Feruglio:2017spp,Acharya:1995ag,Lu:2019vgm,Novichkov:2019sqv,Baur:2019kwi,Dent:2001cc,Giedt:2002ns,Chen:2019ewa}. Here, the couplings retain a modular form and can be expressed as a complex function of modulus $\tau$ \cite{King:2020qaj,Nomura:2019xsb,Feruglio:2017ieh,Behera:2020sfe,Kobayashi:2019xvz}. Once the complex modulus $\tau$ acquires VEV, the symmetry becomes useful to study the neutrino masses and mixing. Hence unlike the usual discrete groups, the importance of scalar fields are somehow being replaced by the Yukawa couplings. Additionally, both fields and couplings transform under a modular group $\Gamma_N$. For various $\Gamma$s, one can infer the isomorphism of different discrete symmetries, for examples $\Gamma_2\simeq S_3$, $\Gamma_3 \simeq A_4$ \cite{Abbas:2020qzc,Wang:2019xbo,Kobayashi:2019gtp} and $\Gamma_4 \simeq S_4$ \cite{Penedo:2018nmg,Liu:2020akv,Gui-JunDing:2019wap} and $\Gamma_5 \simeq A_5$ \citep{Ding:2019zxk,Novichkov:2018nkm}.
   
   The well known permutation group $S_3$ is vastly used for model building purpose due to its simplistic representations \cite{Kubo:2003pd,Meloni:2010aw,Canales:2012dr,Mondragon:2007af}. But this needs atleast three Higgs doublets to explain the experimental results in quark and lepton sectors \cite{Araki:2005ec}. However the modular form can make it more simpler due to the introduction of less scalar fields, where the major role will be played by the Yukawa couplings and there exist very few literature in this direction \cite{Kobayashi:2019rzp}. Also the type III seesaw scenario within $S_3$ symmetry is less frequented as compared to the type I and type II mechanisms. Moreover there exist immense literature on the generic scenario of leptogenesis within the framework of type I and II seesaw but very few studies explore the same in type III case \cite{Davidson:2008bu,Weinberg:1980bf,Pascoli:2006ie,Sierra:2014tqa,Felipe:2013kk,Hambye:2003ka,Senami:2003jn,Lavignac:2015gpa,AristizabalSierra:2012pv,Chen:2009vx}. Therefore in the present work we explore neutrino masses and mixing within the framework of type III seesaw with implication of modular $S_3$ group. Here, the Yukawa couplings transform non-trivially under the $S_3$ symmetry and replace the need for multiple scalar fields. Along with the neutrino mass problem, matter-anti matter asymmetry of the universe remains an attractive question to be proven by experiments but the mystery is till yet unsolved. Thus we try to explore the the generation of lepton asymmetry from the decay of lightest heavy triplet to adequate the explanation of baryon asymmetric universe through leptogenesis phenomena. Unlike the type I seesaw, where the right handed neutrinos being gauge singlets do not have any gauge scattering processes to affect the lepton number density. But the fermion triplets being charged under gauge symmetry leads to gauge interactions, which is effectively contribute to the evolution of their number densities.

   The manuscript is structured as: The section II includes the detail description of model and Lagrangian along with the charged lepton masses. The neutrino masses and mixing within the framework of type III seesaw is discussed in section III. In section IV, we explore the scenario of leptogenesis in detail followed by the solutions of Boltzmann equations. We finally summarize the work in section V.
 
\section{The model}
In the current section, we introduce the particle spectrum and corresponding group charges of the model. The SM particle content is extended with the inclusion of three fermion triplets ($\Sigma_{1,2,3}$) and one scalar singlet ($\rho$). First two fermion triplets are combined to transform as a doublet under the $S_3$ symmetry, where the third triplet transform as a singlet. All the fermion triplets are assigned with a modular weight $-1$. However, the new scalar transform as a singlet under $S_3$ symmetry with assignment of modular weight $-2$. Similarly the first two lepton generations ($L_{1,2}$) transform as a doublets under the $S_3$ symmetry and the third family ($L_3$) remains singlet. The anti-particle states of these leptons are assigned with modular weight of $-1$. The right handed lepton generations transform in a similar manner and their modular weights ($1$) are adjusted to make the Lagrangian invariant. Three Yukawa couplings are introduced to transform as doublet ($y_1(\tau),~y_2(\tau)$) under $S_3$ modular group with a modular weight of 2 and one Yukawa ($y_3 (\tau)$) transform as a singlet with modular weight 4. All the particles and their charges under ${\rm SU(2)_L \otimes U(1)_Y }\otimes S_3$ are provided in Table \ref{partcle content} and \ref{Yukaw couplings}.  

\begin{center} 
\begin{table}[h!]
\centering
\begin{tabular}{|c||c|c|c|c|c|c||c|c|c|c|}\hline\hline  
  & \multicolumn{6}{c||}{Fermions} & \multicolumn{2}{c|}{Scalars}  \\ \hline \hline

& ~($E_{1R}, E_{2R}$)~& $E_{3R}$ & ~($L^c_1, L^c_2$)~& ~$L^c_3$~& ~($\Sigma_{1R}, \Sigma_{2R}$)~& ~$\Sigma_{3R}$~& ~$H$~&~$\rho$ \\ \hline 
${\rm SU(2)_L}$ & 1  & $1$  & $2$  & $2$  & $3$  & $3$  & $2$   & $1$     \\\hline 
${\rm U(1)_Y}$  & $-2$  & $-2$   & $1$      & $1$      & $0$       & $0$       & $1$        & $0$     \\\hline 
$S_3$     & $2$  & $1$  & $2$  & $1$  & $2$  & $1$  & $1$   & $1$  \\ \hline
$k_I$     & $1$       & $1$       & $-1$       & $-1$       & $-1$      & $-1$       & $0$        & $-2$       \\ \hline
\end{tabular}
\caption{Particle content of the model and their charges under $S{\rm U(2)_L\otimes U(1)_Y}\otimes S_3$, where $k_I$ denotes the modular weight.}
\label{partcle content}
\end{table}
\end{center}
\vspace{-2cm}
\begin{center} 
\begin{table}[h!]
\begin{tabular}{|c||c|c|c|c|c|}\hline
{Couplings}  & ~{ $A_4$}~& ~$k_I$~     \\\hline 
{ $(y_1(\tau),~~y_2(\tau))$} & ${\bf 2}$ & ${\bf 2}$      \\\hline
{ $y_3 (\tau)$} & ${\bf 1}$ & ${\bf 4}$      \\\hline
$\lambda_\rho$ & ${\bf 1}$ & ${\bf 8}$ \\\hline
\end{tabular}
\caption{Modular weight of the Yukawa and quartic couplings and their transformation under $S_3$ symmetry.}
\label{Yukaw couplings}
\end{table}
\end{center}
\subsection{Modular Transformation}
A set of linear fractional transformation operates on complex modulus $\tau$ in the upper-half complex plane, forms a modular group $\tilde{\Gamma}$ \cite{Feruglio:2017spp,Kobayashi:2018wkl}.
The transformation leads as following
\begin{equation}\label{eq:tau-SL2Z}
\tau \longrightarrow \gamma\tau= \frac{a\tau + b}{c \tau + d}\ ,~~
{\rm where}~~ a,b,c,d \in \mathbb{Z}~~ {\rm and }~~ ad-bc=1, 
~~ {\rm Im} [\tau]>0 ~ ,
\end{equation}
This is isomorphic to $PSL(2,\mathbb{Z})=SL(2,\mathbb{Z})/\{I,-I\}$.
The modular transformations in $S$ and $T$ diagonal basis are defined as
\begin{eqnarray}
S:\tau \longrightarrow -\frac{1}{\tau}\ , \qquad\qquad
T:\tau \longrightarrow \tau + 1\ ,
\end{eqnarray}

We can have a set of modular groups: $\Gamma(N)~ (N=1,2,3,\dots)$, which are denoted as follows
 \begin{align}
 \begin{aligned}
 \Gamma(N)= \left \{ 
 \begin{pmatrix}
 a & b  \\
 c & d  
 \end{pmatrix} \in SL(2,\mathbb{Z})~ ,
 ~~
 \begin{pmatrix}
  a & b  \\
 c & d  
 \end{pmatrix} =
  \begin{pmatrix}
  1 & 0  \\
  0 & 1  
  \end{pmatrix} ~~({\rm mod} N) \right \}
 \end{aligned} .
 \end{align}
 
For $N=2$, $\tilde\Gamma(2)\equiv \Gamma(2)/\{I,-I\}$, where as for $N>2$ one can define $\tilde\Gamma(N)= \Gamma(N)$. Quotient groups, which come from the finite modular group are defined as $\Gamma_N\equiv \tilde \Gamma/\tilde\Gamma(N)$. The modular groups $\Gamma_N$ with $N=2,3,4,5$ are isomorphic to
$S_3$, $A_4$, $S_4$ and $A_5$ respectively \cite{deAdelhartToorop:2011re}.
The level $N$ modular forms are holomorphic functions $f(\tau)$ and this transforms as following
\begin{equation}
f(\gamma\tau)= (c\tau+d)^k f(\tau)~, ~~ \gamma \in \Gamma(N)~ ,
\end{equation}
here, $k$ is known to be the modular weight.
Since in the present context, we will discuss the modular $S_3$ symmetric group with $N=2$, any field $\phi^{(I)}$ transforms Eq.(\ref{eq:tau-SL2Z})
\begin{equation}
\phi^{(I)} \to (c\tau+d)^{-k_I}\zeta^{(I)}(\gamma)\phi^{(I)},
\end{equation}
here, $\zeta^{(I)}(\gamma)$ denotes the unitary representation matrix of $\gamma\in\Gamma(2)$. The kinetic term for the scalar field is defined as
\begin{equation}
\sum_I\frac{|\partial_\mu\phi^{(I)}|^2}{(-i\tau+i\bar{\tau})^{k_I}} ~,
\label{kinetic}
\end{equation}
Since the Yukawa couplings transform non-trivially under the $S_3$ symmetry and assigned with finite modular weights, they can be expressed in terms of Dedekind eta functions ($\eta$) and their derivatives ($\eta^\prime$) as following \cite{Okada:2019xqk}
\begin{eqnarray}
&& y^{(2)}_1 (\tau) = \frac{i}{4 \pi} \frac{\eta^\prime(\tau /2)}{\eta(\tau /2)} + \frac{\eta^\prime ((\tau+1)/2)}{\eta ((\tau+1)/2)}-\frac{ 8 \eta^\prime (2\tau)}{\eta (2\tau)}, \label{mod_yuk1}\\
&& y^{(2)}_2 (\tau) =\frac{\sqrt{3} i}{4\pi} \left( \frac{\eta^\prime (\tau /2)}{\eta (\tau/2)} - \frac{\eta^\prime ((\tau +1)/2)}{\eta((\tau+1)/2)}\right),\\
&& y^{(4)}_3 (\tau) = \left[(y_1(\tau),~ y_2 (\tau)) \otimes (y_1(\tau),~y_2(\tau))\right]_{1} = y^2_1(\tau) +y^2_2(\tau).  \label{mod_yuk3}
\end{eqnarray}
\subsection{Scalar potential}
Since we have one SM Higgs doublet and a singlet scalar $\rho$ with modular weight 0 and $-2$ respectively, we can write the interaction potential with required Yukawa and quartic coupling along with the free parameters $\alpha^{\prime \prime}$ and $\beta^{\prime \prime}$ as following
\begin{eqnarray}
V= \mu^2_H (H^\dagger H)+\lambda_H(H^\dagger H)^2+y_3 \mu^2_\rho(\rho^\dagger \rho)+\alpha^{\prime \prime}\lambda_\rho (\rho^\dagger \rho)^2 + \beta^{\prime \prime}y_3 (H^\dagger H) (\rho^\dagger \rho).
\end{eqnarray}
The vacuum expectation values of the scalars can be written as $\langle H \rangle = \frac{1}{\sqrt{2}} \begin{pmatrix} 0 \\ v\end{pmatrix} $ and $\langle \rho \rangle =\frac{v_\rho}{ \sqrt{2}}$.
\subsection{Charged lepton masses}
As per the symmetric and modular weight assigned to the fermion doublets, one can write the charged lepton invariant Lagrangian as following
\begin{eqnarray}
\mathcal{L}_l = -y_e \left[\bar{L_1}H E_{1R}+\bar{L_2} H E_{2R}\right]- y_\tau \left[\bar{L_3} H E_{3R}\right]-y_{SB}\left(\bar{L_1}H E_{2R} + \bar{L_2}H E_{1R}\right). \label{cl_lgrng}
\end{eqnarray}
Since the transformation of first two generation leptons as doublet under the $S_3$ symmetry, which leads to degenerate masses for them. Therefore the soft symmetry breaking term $y_{SB}$ is introduced to explain the correct charged lepton masses by finetuning. The mass matrix for the charged lepton can be structured as
\begin{equation}
M_\ell =\frac{v}{\sqrt{2}}\begin{pmatrix}
y_e  & y_{SB} & 0\\
y_{SB} & y_e & 0 \\
0 & 0 & y_\tau
\end{pmatrix}. \label{cl_massmatrix}
\end{equation}
Therefore it is straightforward to obtain the mixing matrix that diagonalizes the charged lepton masses, which can be expressed as following
\begin{equation}
U_{el}=\begin{pmatrix}
\cos{\theta} & -\sin{\theta} & 0\\
\sin{\theta} & \cos{\theta} & 0\\
0 & 0 &1
\end{pmatrix}
\end{equation}
Hence, we can have the mass eigenvalues $m_\mu = \frac{v}{\sqrt{2}}(y_e + y_{SB}) $, $m_e = \frac{v}{\sqrt{2}} (y_e - y_{SB}) $ and $m_\tau = \frac{y_\tau v}{\sqrt{2}}$. The Yukawa couplings and the soft breaking parameter can be adjusted to obtain observed masses of charged leptons.

\section{Type III seesaw Neutrino masses}
The fermion triplets are defined in $\rm SU(2)$ basis and is given by \cite{Hambye:2013jsa,Bandyopadhyay:2009xa}
\begin{eqnarray}
\Sigma_i=\begin{pmatrix}
\frac{{\Sigma_i}^0}{\sqrt{2}} && {\Sigma_i}^{+}\\
 {\Sigma_i}^{-} && -\frac{{\Sigma_i}^0}{\sqrt{2}}\\
\end{pmatrix}.
\end{eqnarray}
The interaction Lagrangian invariant under the $SU(2)\times U(1)_Y \times S_3$ symmetries, which involves the fermion triplets, scalars and lepton doublets can be written as following 
\begin{eqnarray}
\mathcal{L}_\nu &=& -y_1(\tau)\left[\bar{L_1}\Sigma_{2R} \tilde{H}+\bar{L_2}\Sigma_{1R} \tilde{H}\right]\alpha - y_2(\tau)\left[\bar{L_1}\Sigma_{1R} \tilde{H}-\bar{L_2}\Sigma_{2R} \tilde{H}\right]\alpha \nn \\
&& -y_1(\tau)\left[\bar{L_1} \Sigma_{3R} \tilde{H}\right] \gamma -y_2(\tau)\left[\bar{L_2} \Sigma_{3R} \tilde{H}\right]\gamma -y_1(\tau)\left[\bar{L_3} \Sigma_{1R} \tilde{H}\right]\beta \nn \\
&&-y_2(\tau)\left[\bar{L_3} \Sigma_{2R} \tilde{H}\right]\beta -y_3 (\tau) \left[\bar{L_3} \Sigma_{3R} \tilde{H} \frac{\rho}{\Lambda}\right]\alpha^\prime+~{\rm H.c}.
\end{eqnarray} 
Here, $\alpha$, $\beta$, $\gamma$ and $\alpha^\prime$ are the free parameters. Instead of writing $y_i (\tau)$, we use the notation $y_i$ for the following expressions to avoid lengthy conventions. Now we can construct the flavor structure of Dirac mass matrix for neutrinos as follows
\begin{equation}
M_D=\frac{v}{\sqrt{2}}\begin{pmatrix}
\alpha y_2 & \alpha y_1 & \gamma y_1\\
\alpha y_1 & - \alpha y_2 & \gamma y_2\\
\beta y_1 & \beta y_2 & \alpha^\prime y_3 \frac{v_\rho}{\Lambda}
\end{pmatrix}.
\end{equation}
The Lagrangian for the fermion triplet involves the kinetic and mass terms is given by
\begin{eqnarray}
\mathcal{L}_\Sigma &=& -i {\rm Tr} \left[\bar{\Sigma_{iR}}\gamma^\mu D_\mu \Sigma_{iR} \right] -\frac{1}{2}  {\rm Tr}\left[y_3 \overline{\Sigma^c_{3R}} \Sigma_{3R} \frac{\rho}{\Lambda}\right]M^\prime_0 \nn \\
&&-\frac{1}{2} {\rm Tr}\left[y_1 (\overline{\Sigma^c_{1R}} \Sigma_{2R} + \overline{\Sigma^c_{2R}} \Sigma_{1R}) + y_2 (\overline{\Sigma^c_{1R}} \Sigma_{1R} - \overline{\Sigma^c_{2R}} \Sigma_{2R}) \right]M_0.
\end{eqnarray}
Here, $M_0$ and $M^\prime_0$ are the free mass parameters. Thus the mass matrix for the fermion triplets can be constructed as following
\begin{equation}
M_\Sigma = \begin{pmatrix}
M_0 y_2  & M_0 y_1  & 0\\
M_0 y_1 & - M_0 y_2 & 0\\
0 & 0 & M^\prime_0 y_3 \frac{v_\rho}{\Lambda}
\end{pmatrix}.\label{tripletmass}
\end{equation}
The small Majorana mass matrix for the neutrinos can be written as following
\begin{eqnarray}
\mathcal{M}_\nu &&= M_D M^{-1}_{\Sigma} M^T_D \nn \\
&&=\frac{v^2}{2}\begin{pmatrix}
\frac{\alpha^2 y_2}{M_0} +\frac{\gamma^2 \Lambda y^2_1}{M^\prime_0 v_\rho y_3} && y_1 \left(\frac{\alpha^2}{M_0}+ \frac{\gamma^2 \Lambda y_2}{M^\prime_0 v_\rho y_3}\right) && y_1 \left(\frac{\alpha^\prime \gamma}{M^\prime_0}+\frac{\alpha \beta}{M_0}\right) \\
y_1 \left(\frac{\alpha^2}{M_0}+ \frac{\gamma^2 \Lambda y_2}{M^\prime_0 v_\rho y_3}\right) && y_2 \left(-\frac{\alpha^2}{M_0}+ \frac{\gamma^2 \Lambda y_2}{M^\prime_0 v_\rho y_3}\right) &&  y_2 \left(\frac{\alpha^\prime \gamma}{M^\prime_0}+\frac{\alpha \beta}{M_0}\right)\\
y_1 \left(\frac{\alpha^\prime \gamma}{M^\prime_0}+\frac{\alpha \beta}{M_0}\right) && y_2 \left(\frac{\alpha^\prime \gamma}{M^\prime_0}+\frac{\alpha \beta}{M_0}\right) && -\frac{\beta^2 y_2 (y^2_2 -3 y^2_1)}{M_0 (y^2_1 + y^2_2)} + \frac{{\alpha^\prime}^2 y_3 v_\rho }{M^\prime_0 \Lambda} 
\end{pmatrix}. \label{nmassmatrx}
\end{eqnarray}
The above mass matrix is diagonalized numerically and the corresponding eigenvector matrix ($U_\nu$) is obtained and the mixing parameters are discussed in detail with the standard convention of neutrino mixing matrix $U_{\rm PMNS}= U^\dagger_{el} U_{\nu}$, in the numerical analysis section.\\
Now, focusing on the diagonalization of mass matrix for fermion triplets in Eq.\eqref{tripletmass}, we can have the eigenvector matrix as following
\begin{eqnarray}
U_R=\begin{pmatrix}
\frac{u_-}{\sqrt{\rm N_-}} & \frac{u_+}{\sqrt{\rm N_+}} & 0\\
\frac{1}{\sqrt{\rm N_-}} & \frac{1}{\sqrt{\rm N_+}} & 0\\
0 & 0 & 1
\end{pmatrix},
\end{eqnarray}
here,
\begin{equation}
u_{\pm}= \left[\frac{y_2}{y_1} \pm \sqrt{1+\left(\frac{y_2}{y_1}\right)^2}\right], ~~ {\rm N_\pm}= 1+(u_\pm)^2.
\end{equation}
The mass eigenvalues are obtained upon diagonalization $M^D_\Sigma = U_R M_\Sigma U^T_R$ and are given by
\begin{equation}
|M_{\Sigma_1}| =|M_{\Sigma_2}| = |\left(\sqrt{y^2_1 + y^2_2}\right) | M_0, ~~ |M_{\Sigma_3}| = |y_3| M^\prime_0 \frac{v_\rho}{\Lambda}.
\end{equation}
\subsection{Numerical Analysis}
To explore the numerical analysis for this model, we have considered the  following $3\sigma$ observed limit of neutrino oscillation parameters ~\cite{Esteban:2018azc}
\begin{align}
&{\rm NO}: \Delta m^2_{\rm atm}=[2.431, 2.622]\times 10^{-3}\ {\rm eV}^2,\
\Delta m^2_{\rm sol}=[6.79, 8.01]\times 10^{-5}\ {\rm eV}^2, \nn \\
&\sin^2\theta_{13}=[0.02044, 0.02437],\ 
\sin^2\theta_{23}=[0.428, 0.624],\ 
\sin^2\theta_{12}=[0.275, 0.350].\nn\\ \label{eq:mix}
\end{align} 

We randomly vary the model parameters within the following range and constrained them from the experimentally observed $3\sigma$ limit of neutrino oscillation data.
\begin{align}
&{\rm Re}[\tau],~{\rm Im}[\tau] \in [1,2],~~ { \alpha,\gamma } \in ~[0.005,0.01],~~{\beta} \in ~[0.02,0.06],~~{\alpha^\prime} \in ~[0.1,1], \nn \\
& \quad M_0  \in [10^2,5\times 10^4],\quad M^\prime_0 \in[5\times 10^2,10^6],\quad \frac{v_\rho}{\Lambda} =0.1.
\end{align}

\begin{figure}[h!]
\begin{center}
\includegraphics[height=45mm,width=60mm]{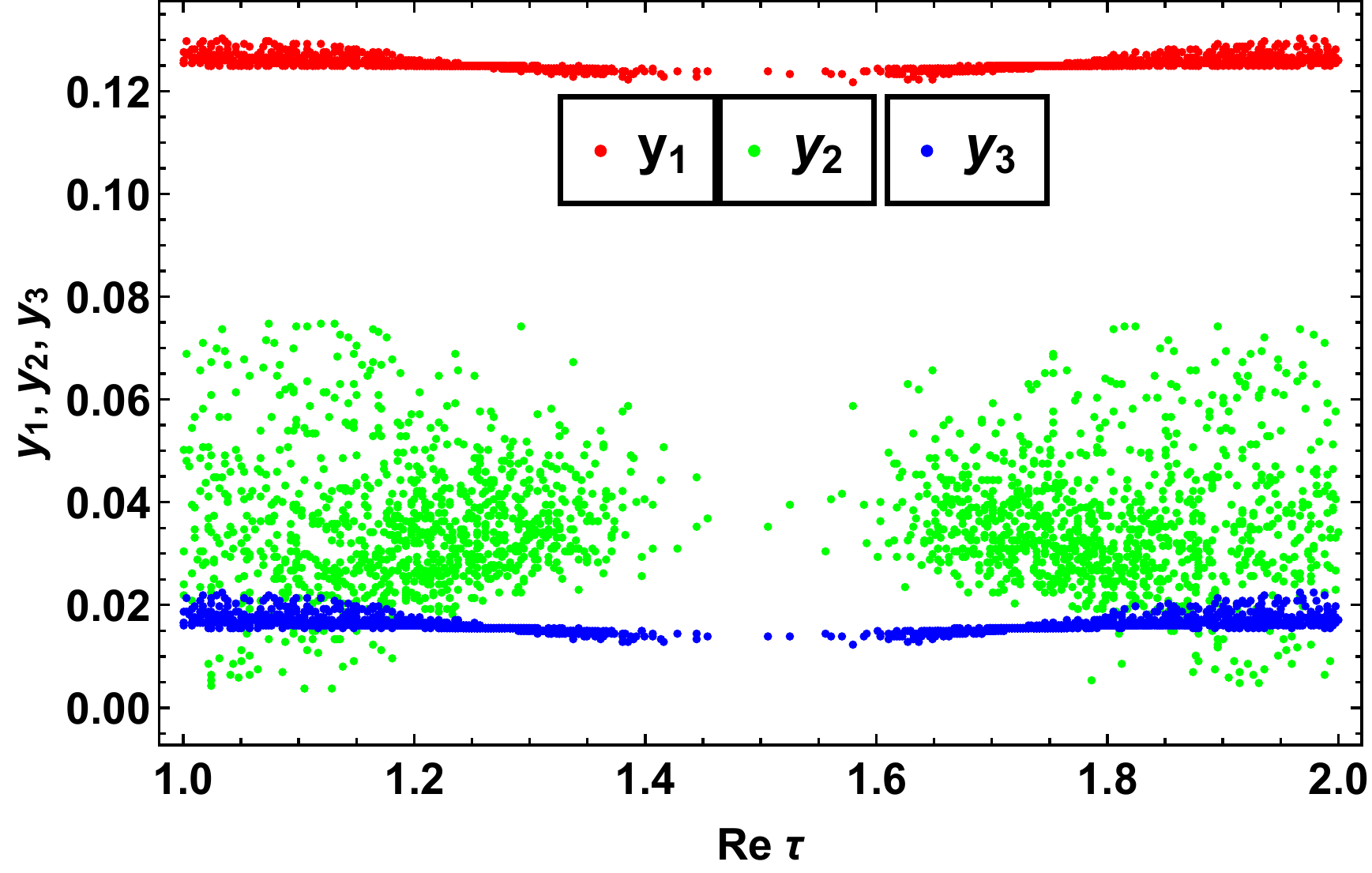}~~~~
\includegraphics[height=45mm,width=60mm]{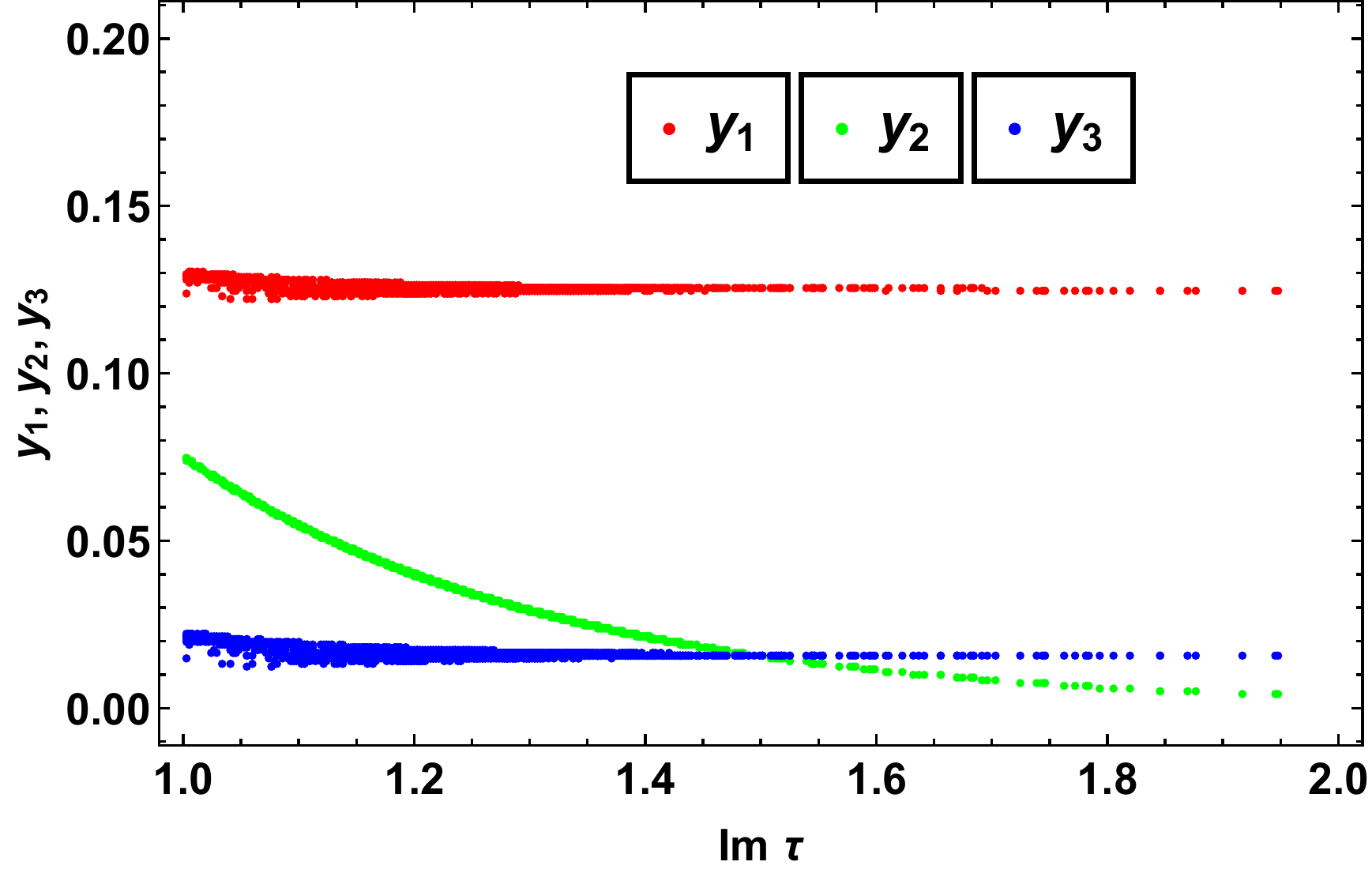}
\caption{Left(Right) panel represents the variation of modular Yukawa couplings with real (imaginary) component of complex modulus $\tau$.}\label{Yuk_tau}
\end{center}
\end{figure}

\begin{figure}[h!]
\begin{center}
\includegraphics[height=45mm,width=60mm]{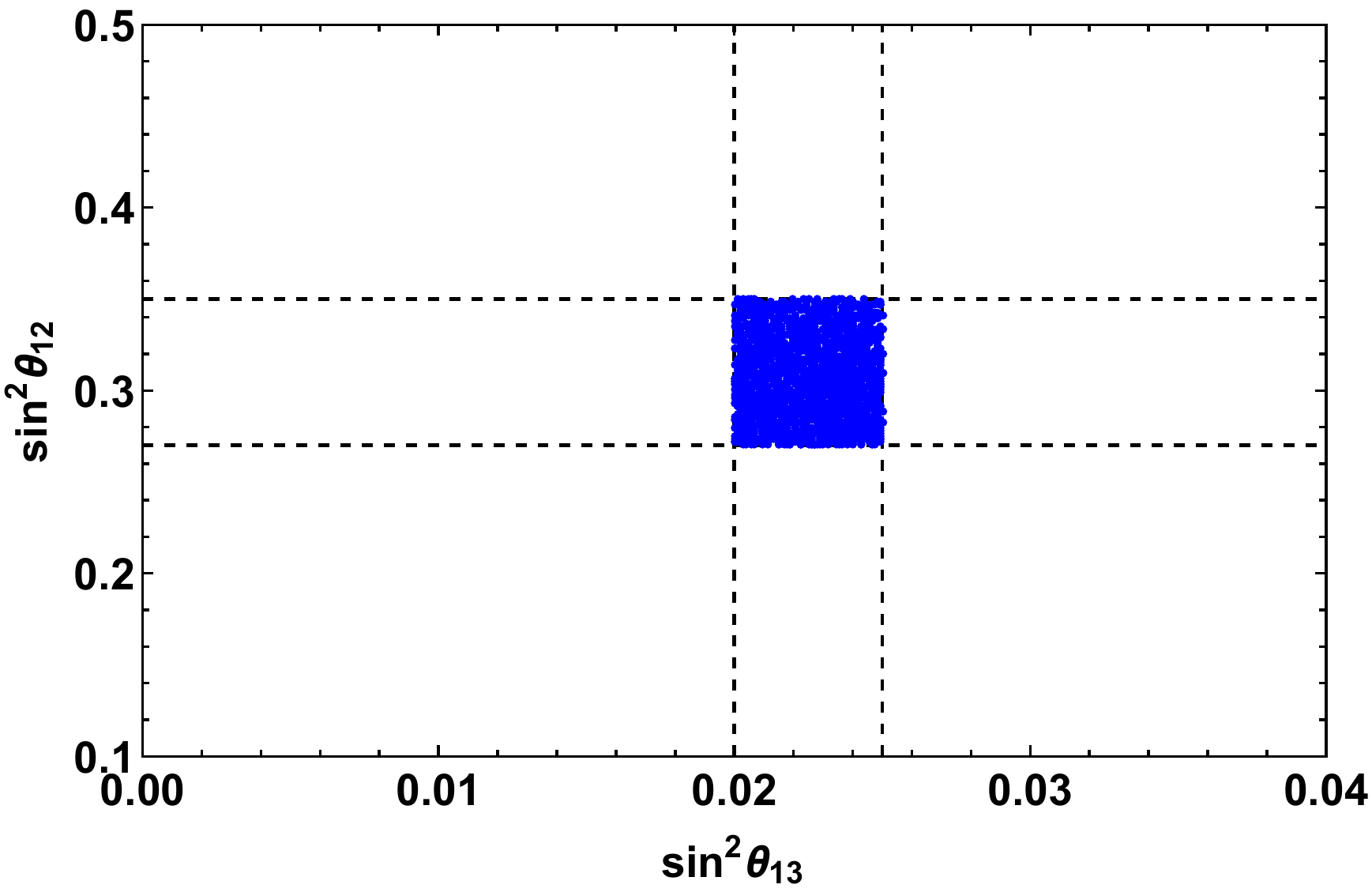}~~~~
\includegraphics[height=45mm,width=60mm]{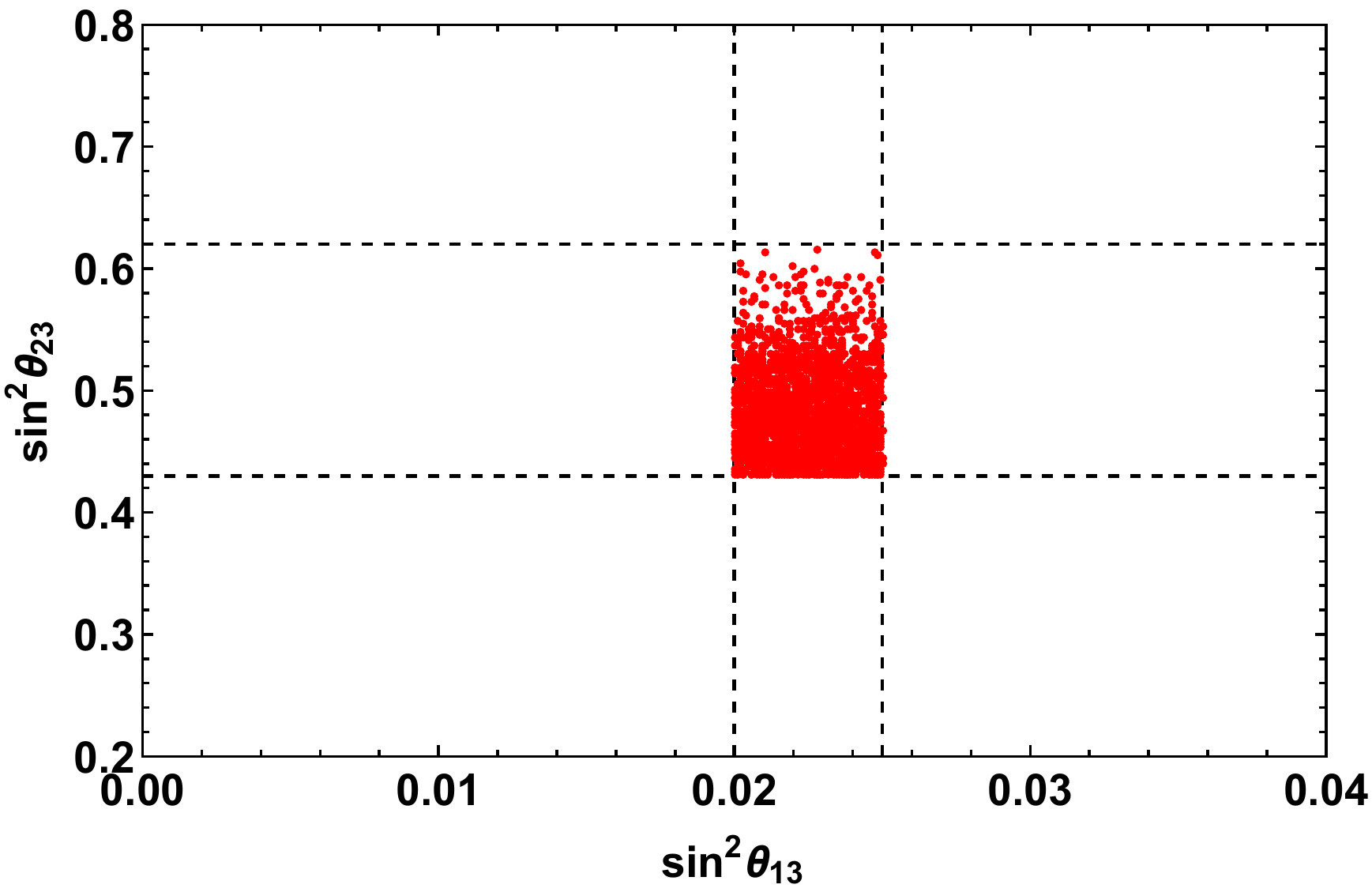}\\
\includegraphics[height=45mm,width=60mm]{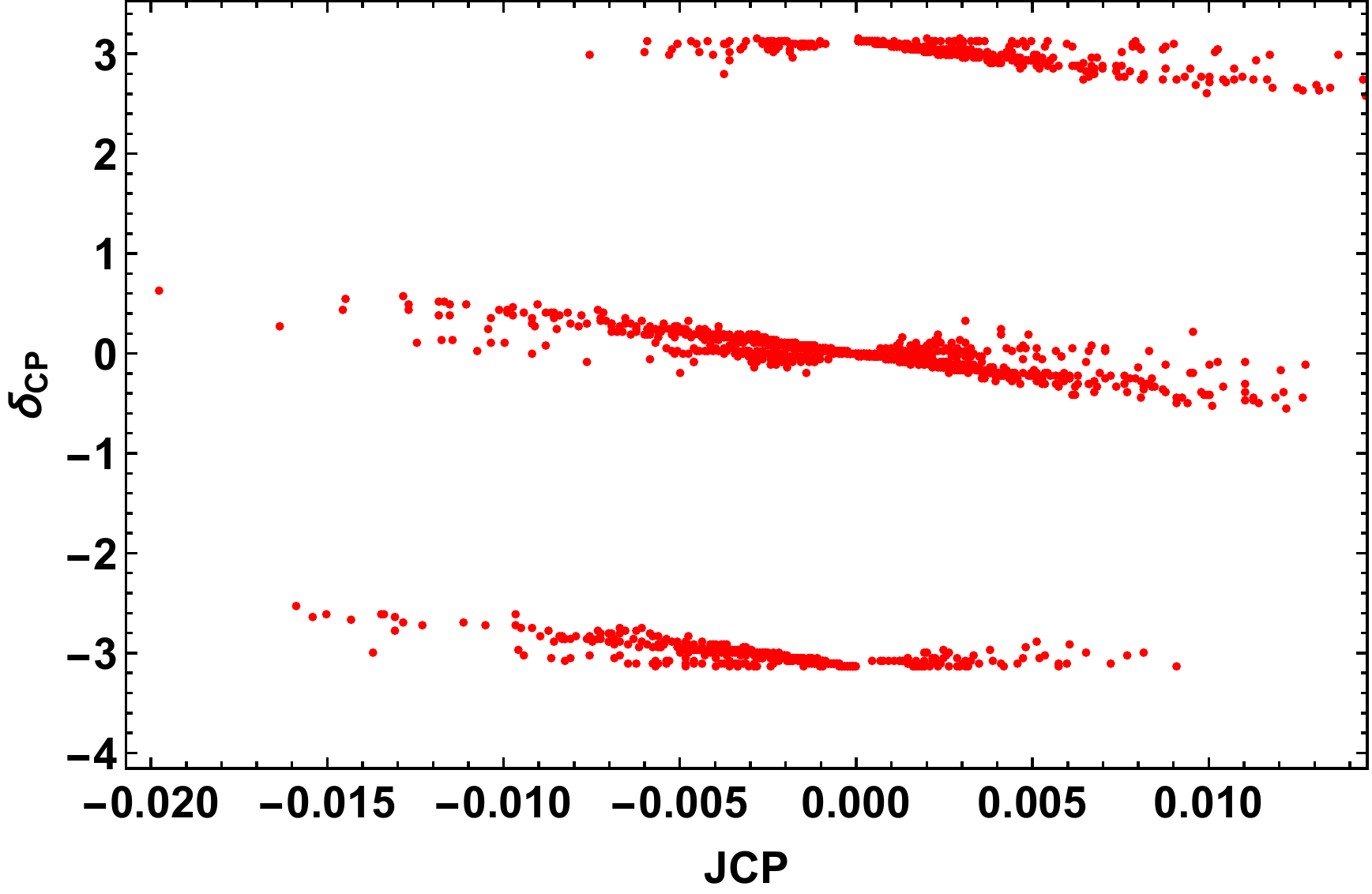}
\caption{Top left panel displays the correlation of reactor mixing angle ($\theta_{13}$) with the solar mixing angle ($\theta_{12}$) and right panel represents the variation of same with the atmospheric mixing angle ($\theta_{23}$) within the $3\sigma$ observed limit. Down panel shows the variation of Dirac rephasing invariant with the Dirac CP violating phase.}\label{angles_CP}
\end{center}
\end{figure}

\begin{figure}[h!]
\begin{center}
\includegraphics[height=45mm,width=60mm]{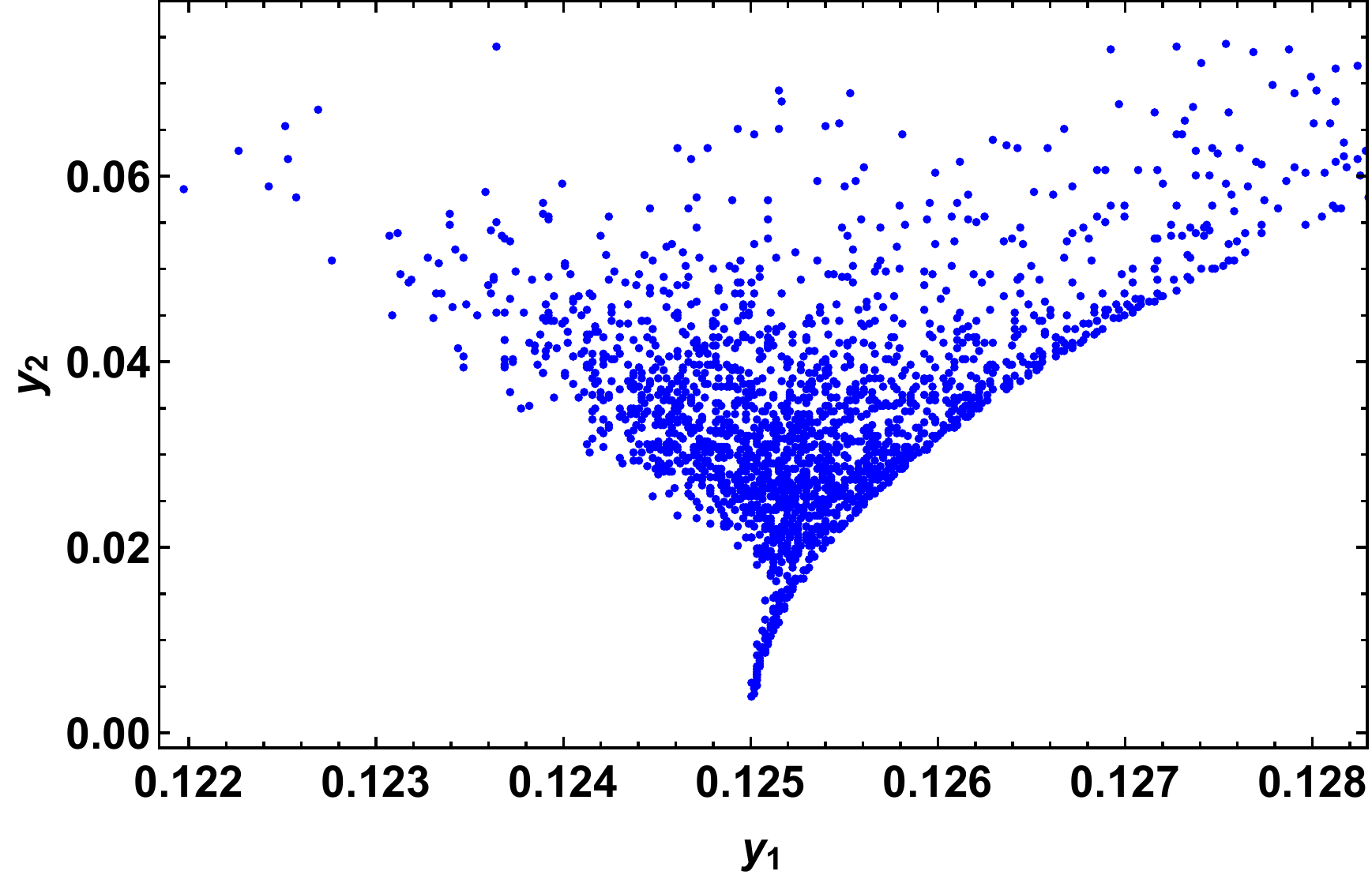}~~~~
\includegraphics[height=45mm,width=60mm]{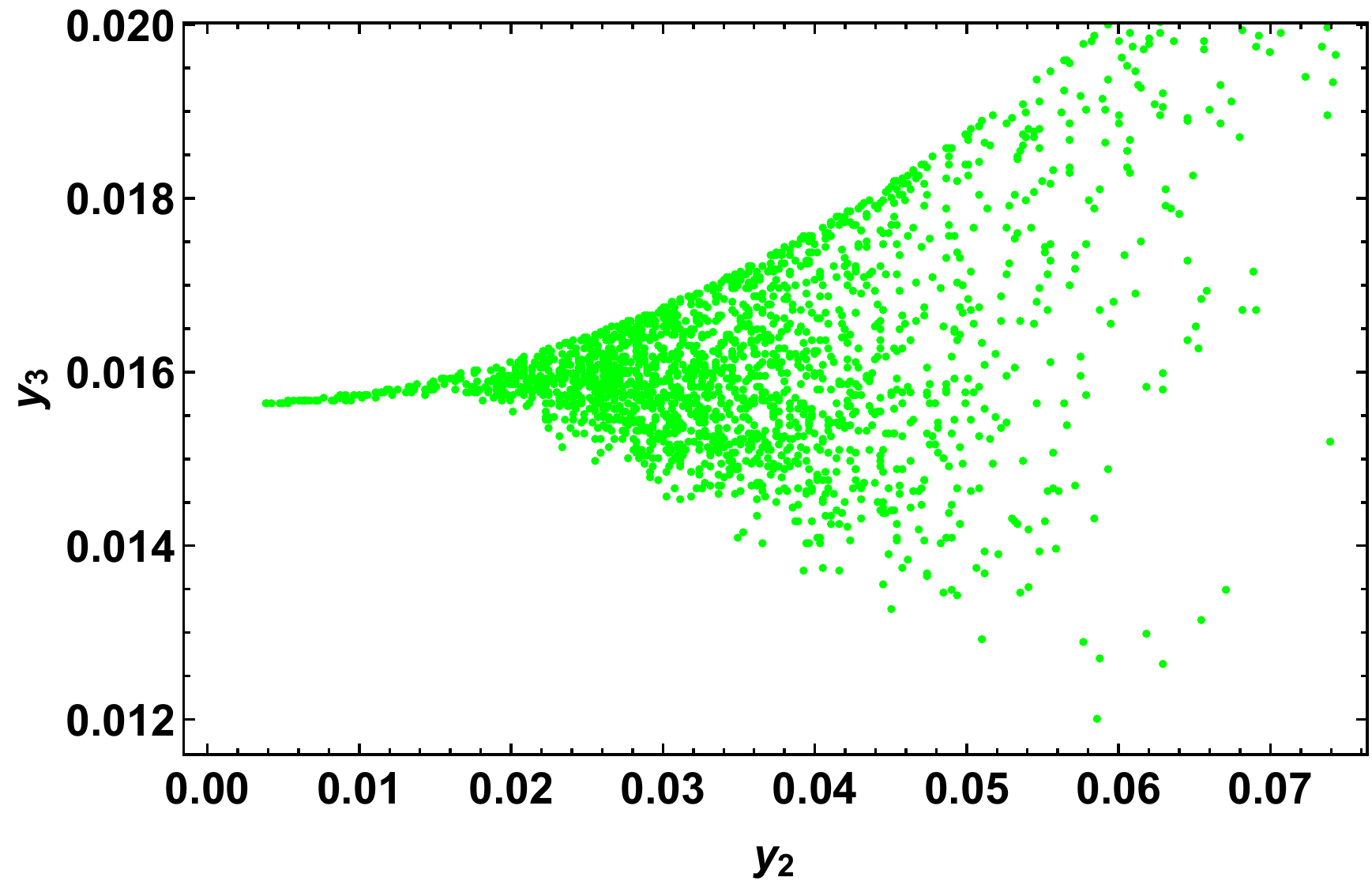}\\
\vspace{3mm}
\includegraphics[height=45mm,width=60mm]{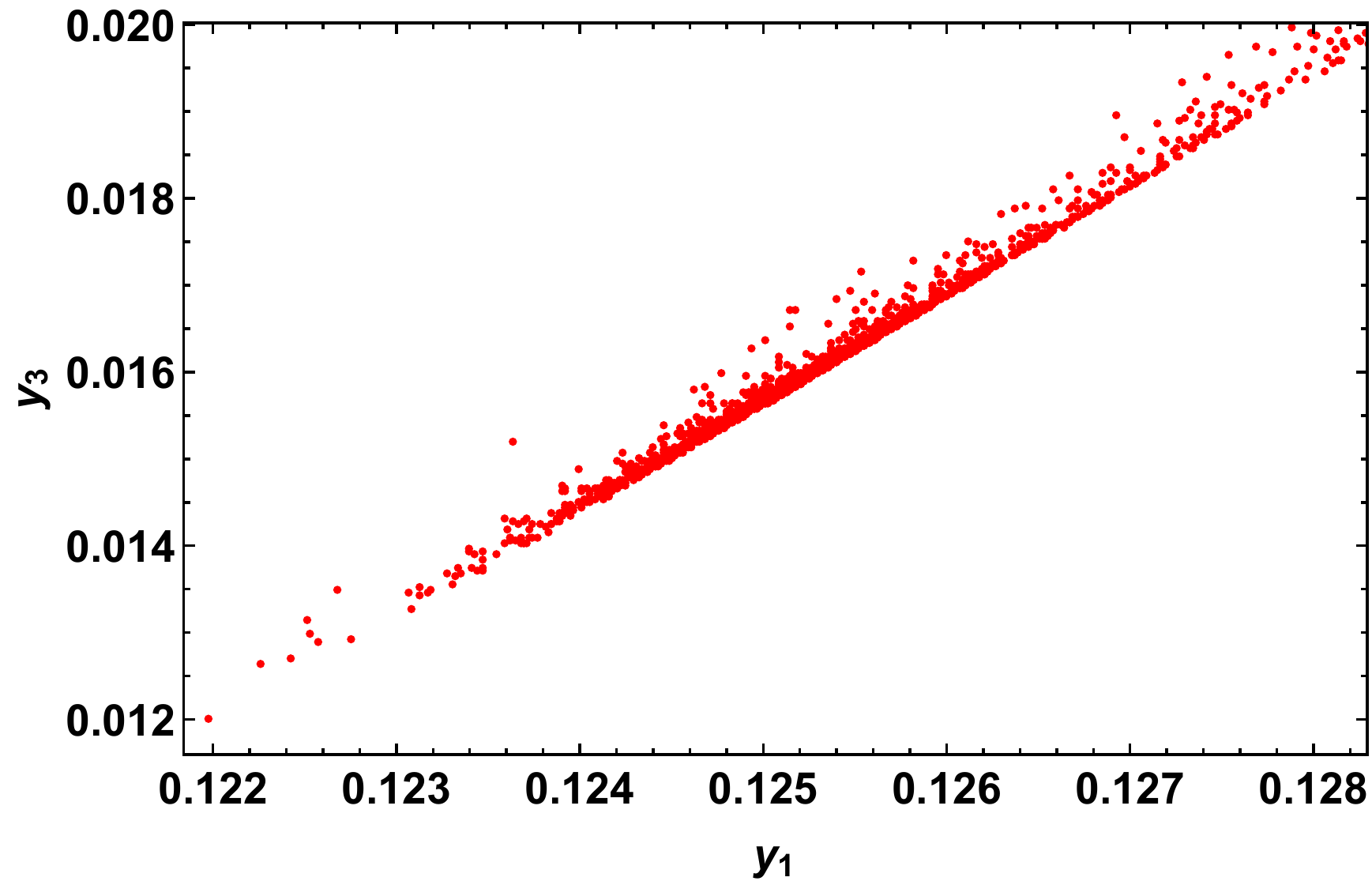}~~~~
\includegraphics[height=45mm,width=60mm]{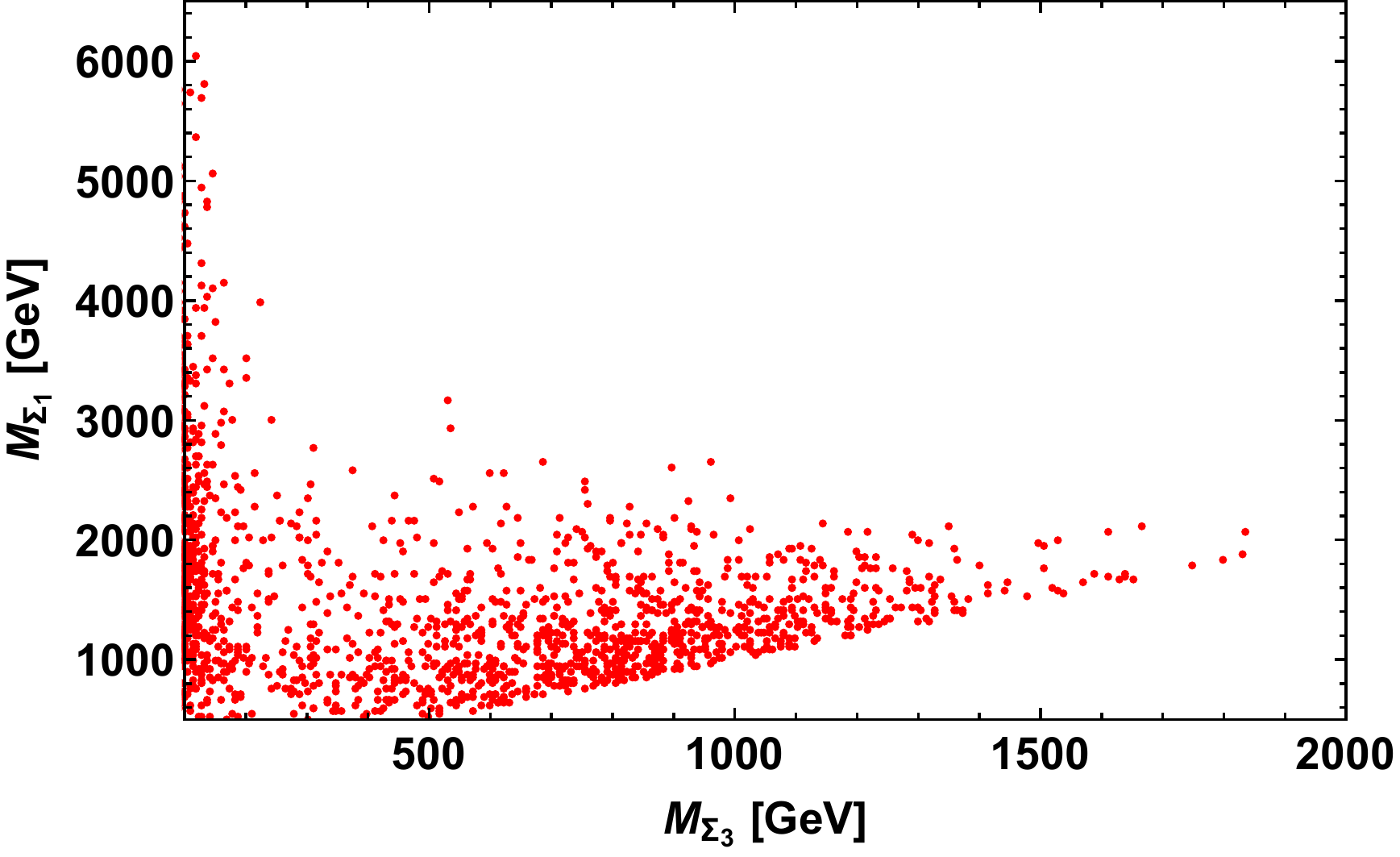}
\caption{Left and right top panel represent the variation of Yukawa couplings $y_1$ with $y_2$ and $y_2$ with $y_3$ respectively. Down left and right panel display the variation of $y_1$ with $y_3$ and correlation of $M_{\Sigma_3}$ with $M_{\Sigma_1}$ respectively.}\label{Yuk_Mr}
\end{center}
\end{figure}

The input values for the model parameters are randomly varied within the above mentioned ranges and their allowed regions are obtained by constraining from the $3\sigma$ observed values of the neutrino oscillation parameters and also the observed neutrino mass bound ($\sum m_{\nu_i} < 0.12$ eV) \cite{Aghanim:2018eyx}.
We consider the complex modulus $\tau$ to vary 1\ $\lesssim\ $Re$[\tau]\lesssim$\ 2 and  1\ $\lesssim\ $Im$[\tau]\lesssim$\ 2 for normal ordering of neutrino masses.
Thus we found the modular Yukawa couplings, which depend on $\tau$ by the relation defined in Eq.\eqref{mod_yuk1} to \eqref{mod_yuk3}, vary within the region $0.12 \lesssim\ $$y_1$$(\tau)\lesssim 0.14$, $ 0 \lesssim\ $$y_2$$(\tau)\lesssim 0.08$ and $  0.01 \lesssim\ $$y_3$$(\tau)\lesssim 0.03 $. The variation of these couplings with the real and imaginary part of the complex modulus are represented in the left and right panel of Fig.\ref{Yuk_tau} respectively. In the top left and right panels of Fig.\ref{angles_CP}, we have represented the variation of mixing angle $\theta_{13}$ with $\theta_{12}$ and $\theta_{23}$ respectively, within their $3\sigma$ observed values. In the down panel, we have shown the correlation of rephasing invariant (JCP) with the Dirac CP violating phase, which is found to lie within the range $\delta_{CP} \in [-0.06,0.06]$ and $[\pm 2.6, \pm 3.14]$ rad.  The correlation of modular Yukawa couplings $y_1$ with $y_2$ and $y_2$ with $y_3$ are displayed in top left and right panel of Fig.\ref{Yuk_Mr} respectively, however in the down left and right panel, we have shown the correlation of $y_1$ with $y_3$ and the variation of Majorana mass for the fermion triplet $M_{\Sigma_1}$ with the lightest heavy triplet mass $M_{\Sigma_3}$.

\section{Leptogenesis}
The well known phenomena of leptogenesis is found to be widely used in the literature due to its simplest formalism. Instead of generating baryon asymmetry directly, one can generate the asymmetry in the lepton sector, which can partially be stored in to the baryon sector during the sphaleron transition \cite{Buchmuller:2004nz,Asaka:2018hyk,Abada:2018oly}. The Davidson Ibara bound on right-handed neutrino mass in case of type I seesaw to be greater than $10^9$ GeV, which is very difficult to be tested in colliders. Thus bringing down the scale of leptogenesis as low as TeV through resonance enhancement, is proven to be an attractive scenario, which may opens up exciting options in the future experiments \cite{Pilaftsis:2003gt,Iso:2010mv,Adhikary:2014qba,Pilaftsis:2005rv}. Obtaining a finite CP asymmetry from the decay of right-handed neutrinos within the simplistic framework of type I seesaw is well explored, however leptogenesis with type III seesaw is less frequented in the literature \cite{Albright:2003xb,Goswami:2018jar,Chen:2009vx}. Thus in the present context, we focus on the generation of asymmetry from the decay of lightest heavy fermion triplet at TeV scale. Since in this model, two of the fermion triplets belongs to the doublet representation of $S_3$ have exactly same masses ($|M_{\Sigma_1}| = |M_{\Sigma_2}|$), we consider the resonance enhancement in the self energy, provided by the condition $|M_{\Sigma_3}| \simeq |M_{\Sigma_1}|$. Since the component of the triplet are having same masses and equal decay width, the CP asymmetry reduces three times than type I case. 
The tree level decay width for the fermion triplet is given by \cite{Franceschini:2008pz}
\begin{equation}
\Gamma_{\Sigma}=\Gamma(\Sigma^0 \rightarrow LH)+\Gamma(\Sigma^0 \rightarrow \bar{L}\bar{H})=\frac{1}{8\pi} M_{\Sigma_i} (\tilde{Y_\Sigma}^\dagger \tilde{Y_\Sigma})_{ii}.
\end{equation}
Here,
\begin{eqnarray}
Y_\Sigma = \begin{pmatrix}
\alpha y_2 & \alpha y_1 & \gamma y_1\\
\alpha y_1 & - \alpha y_2 & \gamma y_2\\
\beta y_1 & \beta y_2 & \alpha^\prime y_3 \frac{v_\rho}{\Lambda}
\end{pmatrix}~ {\rm and}~\tilde{Y_\Sigma}=Y_\Sigma U_{el} U_R.
\end{eqnarray}
\begin{figure}[h!]
\begin{center}
\includegraphics[height=22mm,width=37mm]{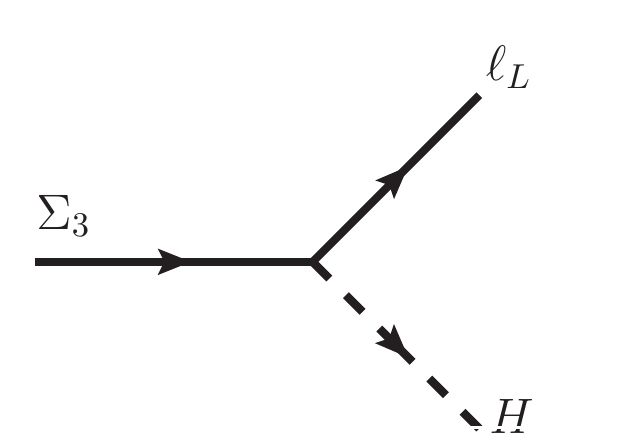}~~~~
\includegraphics[height=22mm,width=37mm]{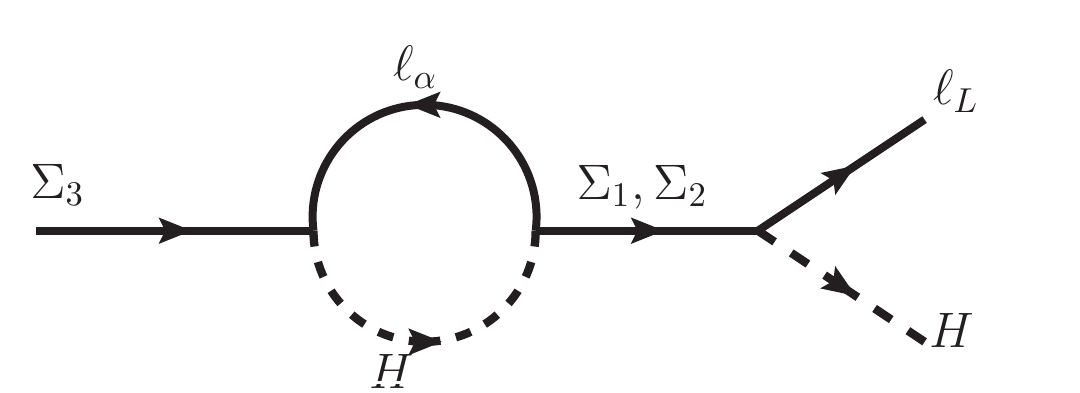}
\includegraphics[height=22mm,width=37mm]{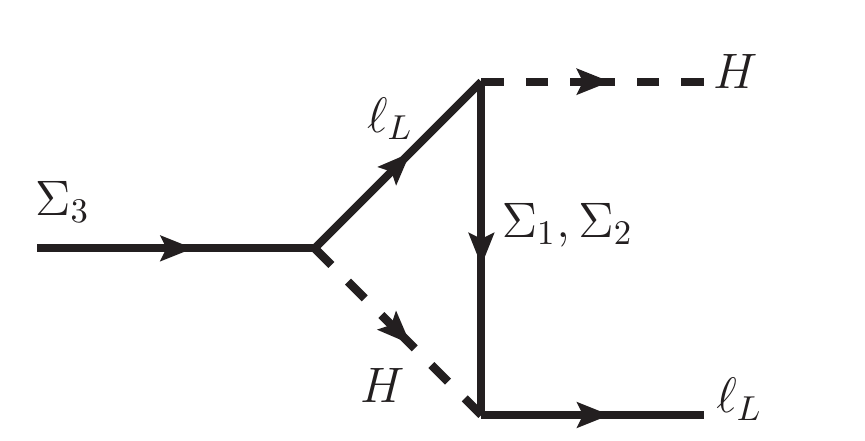}
\caption{Tree and one loop Feynman diagrams for the decay of heavy fermion triplet.}\label{feyndiag}
\end{center}
\end{figure}
The decay width for the charged components of the triplets can be written in a similar way. Unlike the type II case, there is no asymmetry in particle and antiparticle since it does not have several decay modes rather it decays only through Yukawa interaction. The general expression for CP asymmetry can be written as following \cite{Pilaftsis:1997jf,Pilaftsis:2003gt,Hambye:2012fh}
\begin{eqnarray}
\epsilon^\Sigma_{CP} = -\sum_j \frac{3}{2} \frac{M_{\Sigma_i}}{M_{\Sigma_j}} \frac{\Gamma_{\Sigma_i}}{M_{\Sigma_j}} \frac{V-2S}{3} \frac{{\rm Im}\left(\tilde{Y_\Sigma} \tilde{ Y_\Sigma}^
\dagger\right)^2_{ij}}{\left(\tilde{Y_\Sigma} \tilde{Y_\Sigma}^\dagger\right)_{ii} \left(\tilde{Y_\Sigma} \tilde{Y_\Sigma}^\dagger\right)_{jj}}.
\end{eqnarray}
Here V and S are the vertex and self energy contribution respectively. These are expressed as follows
\begin{eqnarray}
&& S=\frac{M^2_{\Sigma_j} \Delta M^2_{ij}}{(\Delta M^2_{ij})^2+M^2_{\Sigma_i} \Gamma^2_{\Sigma_j}},~{\rm with}~ \Delta M_{ij} = M_{\Sigma_j} -M_{\Sigma_i},\\
&& V= \frac{2M^2_{\Sigma_j}}{M^2_{\Sigma_i}}\left[\left(1+\frac{M^2_{\Sigma_j}}{M^2_{\Sigma_i}}\right) {\rm log} \left(1+\frac{M^2_{\Sigma_j}}{M^2_{\Sigma_i}}\right)-1\right]. \label{cp_self}
\end{eqnarray}

\begin{figure}[h!]
\begin{center}
\includegraphics[height=45mm,width=53mm]{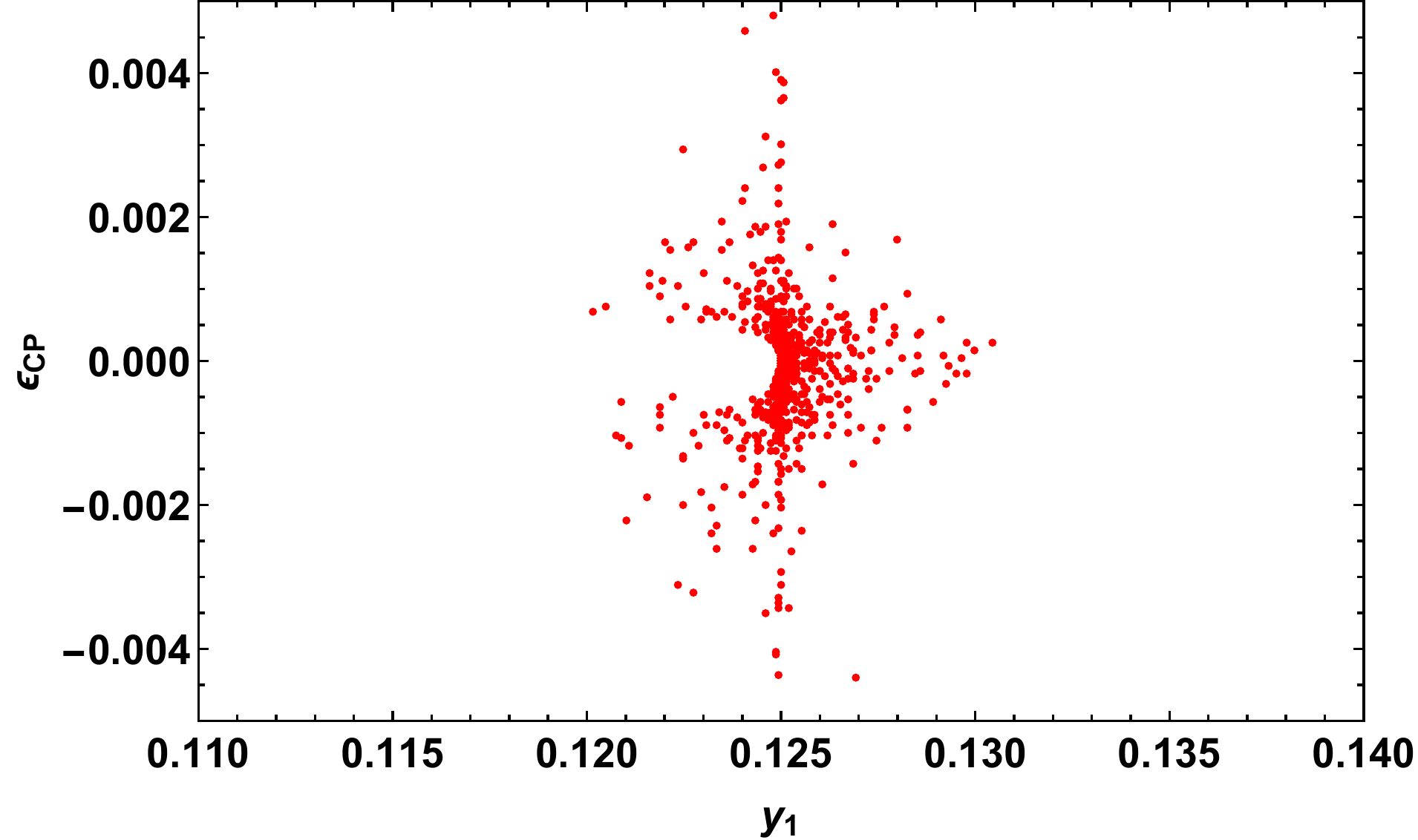}~
\includegraphics[height=45mm,width=53mm]{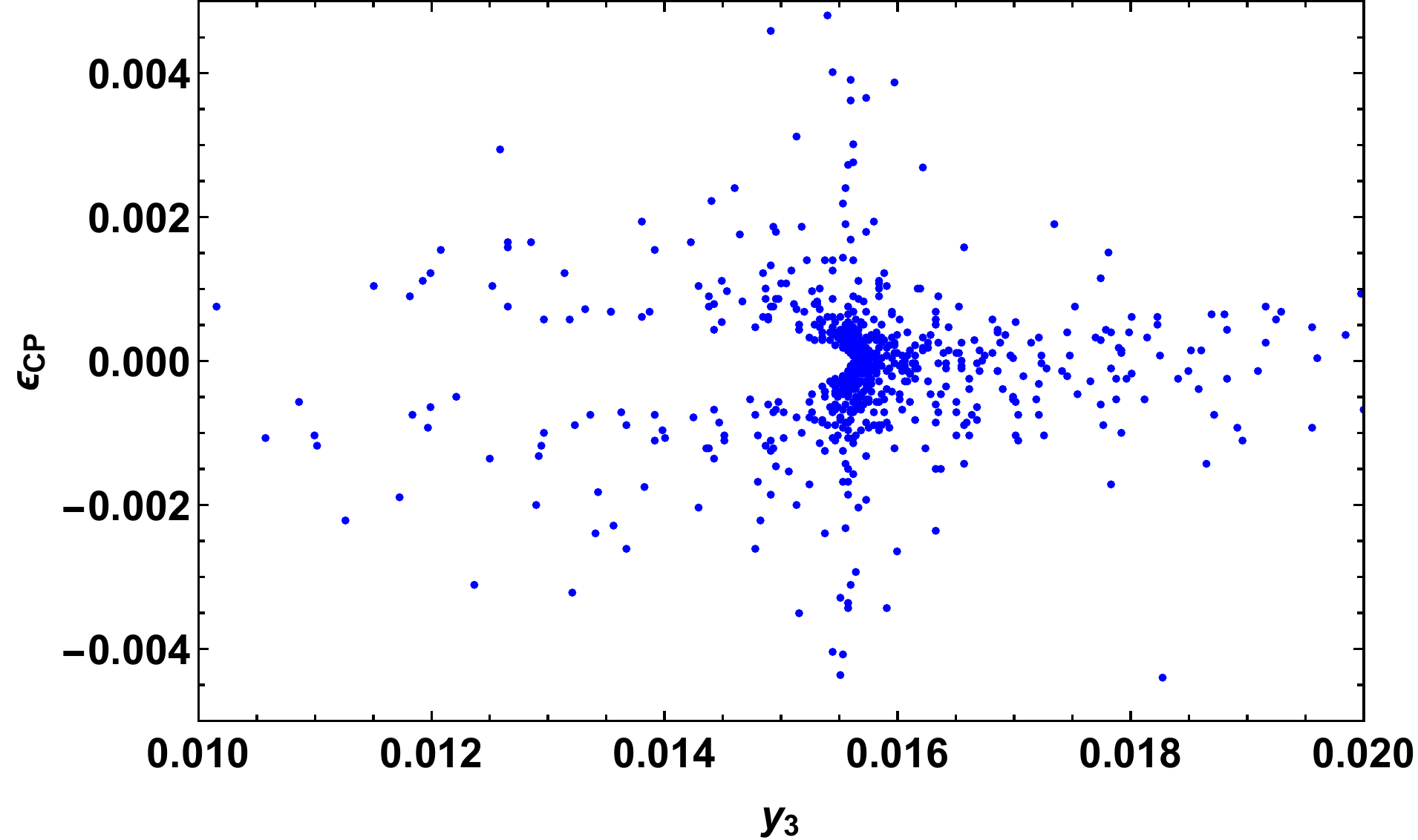}~
\includegraphics[height=45mm,width=53mm]{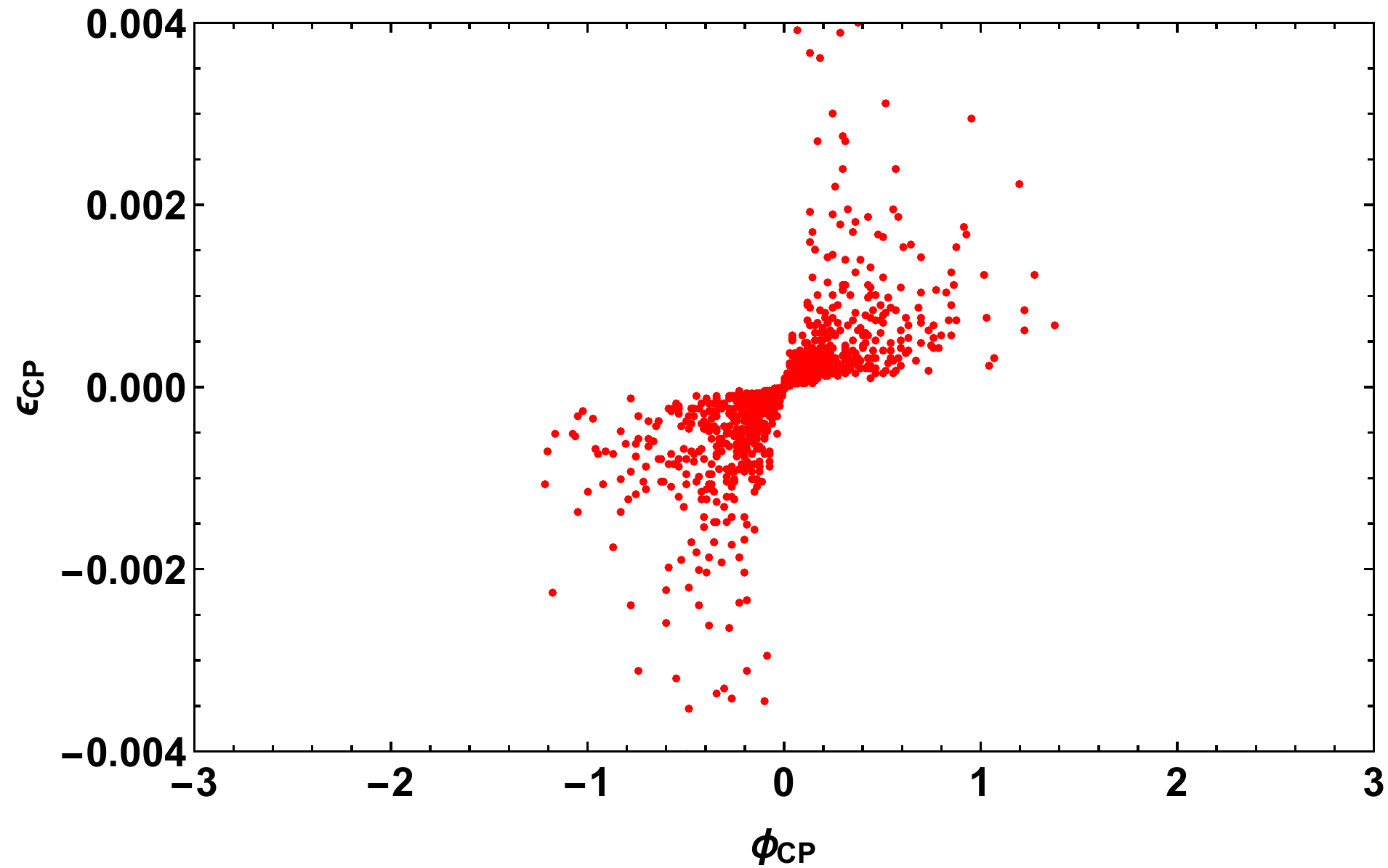}
\caption{Variation of modular Yukawa couplings $y_1$ and $y_3$ with the CP asymmetry parameter is represented in the left and middle panel. The extreme right panel shows the correlation of CP violating phase with the CP asymmetry. }\label{Yuk_CP}
\end{center}
\end{figure}

The Feynman diagrams, those finitely contribute to the CP asymmetry are provided in Fig.\ref{feyndiag}. Since for a TeV scale heavy particle, the CP asymmetry comes out to be small and not enough to generate the required asymmetry, the scenario of resonant leptogenesis is preferred. If we consider the mass difference between the heavy states to be comparable with the decay width, one can clearly infer from Eq.\eqref{cp_self} a resonantly enhanced self energy contribution, i.e $M_{\Sigma_j} - M_{\Sigma_i} \approx \frac{\Gamma_\Sigma}{2}$, which leads an enhancement of S value upto $1/2$ with an almost similar order Yukawa couplings. Therefore we can safely neglect the contribution from the vertex diagram.
We vary the modular Yukawa couplings ($y_1$, $y_3$), which satisfy the neutrino oscillation constraints, with the CP asymmetry parameter and found its value to be order $\approx \mathcal{O} (10^{-3})$. The variation of CP violating phase with the CP parameter is displayed in the right panel of Fig.\ref{Yuk_CP}.

\subsection{Boltzmann Equations}
\begin{figure}[h!]
\begin{center}
\includegraphics[height=50mm,width=65mm]{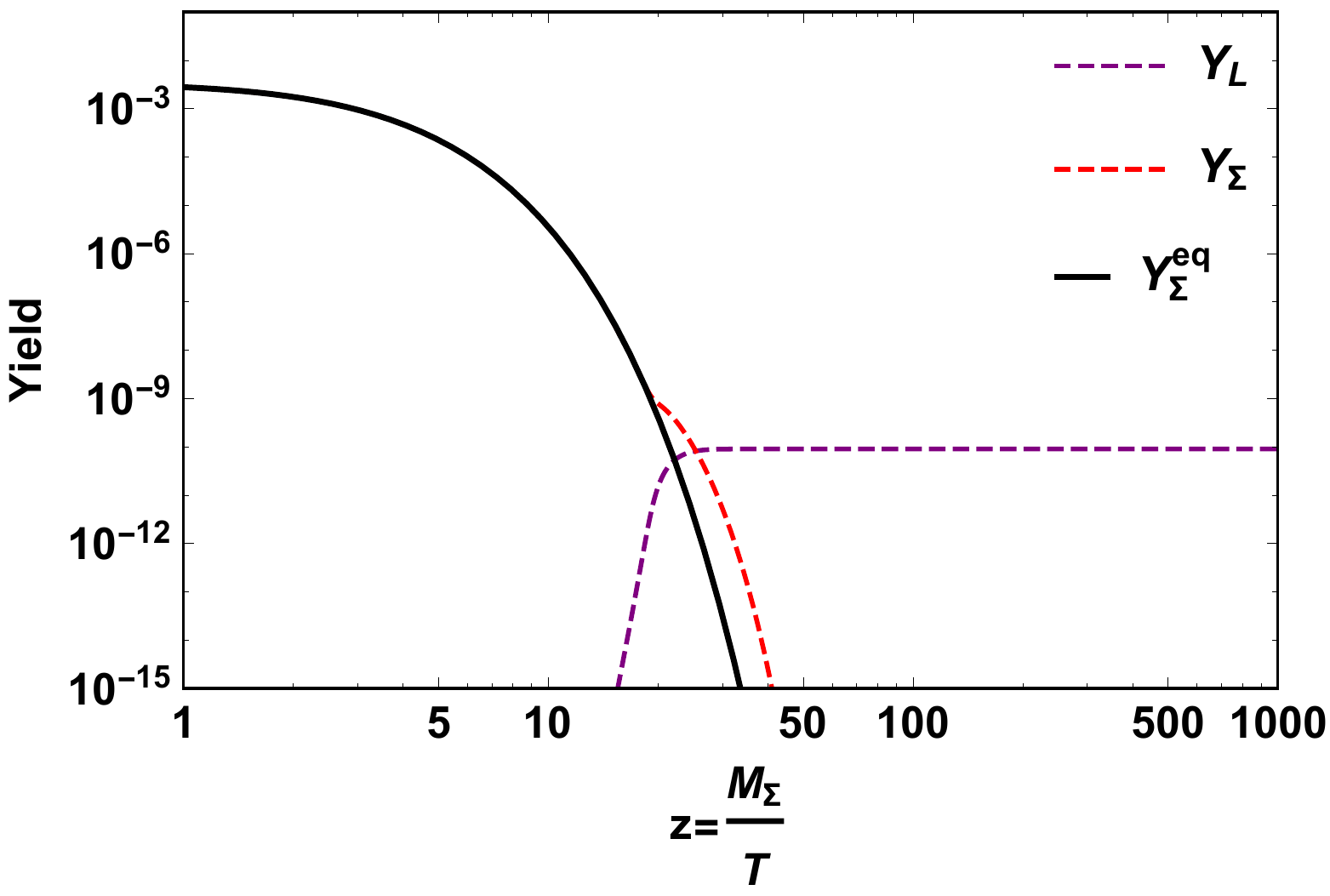}~~~~
\includegraphics[height=50mm,width=65mm]{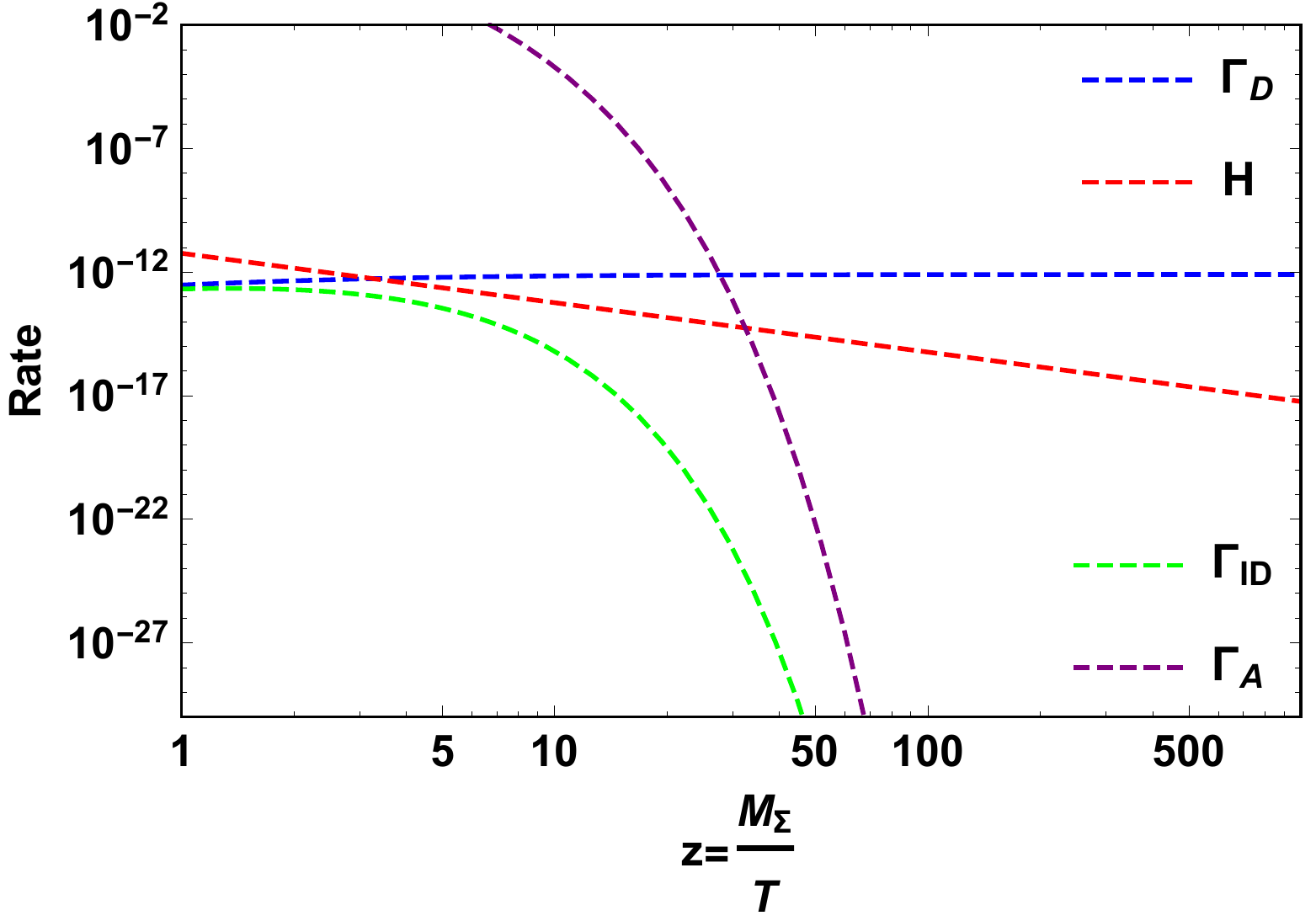}
\caption{Solution to Boltzmann equations to find the evolution of particle number densities.}\label{Beqns}
\end{center}
\end{figure}

The evolution of particle number densities are governed by the dynamics of relevant Boltzmann equations. Sakharov conditions \cite{Sakharov:1967dj} demands the decaying fermion to remain out of equilibrium to generate the asymmetry in lepton sector. One need to compare the Hubble expansion with the decay rate to satisfy this condition, which is given follows. 
\begin{equation}
K_{\Sigma_i} = \frac{\Gamma_{\Sigma_i}}{H(T=M_{\Sigma_i})}.
\end{equation}
Here, $H = \frac{1.67 \sqrt{g_\star}~ T^2 }{M_{\rm Pl}}$, with $g_{\star} = 106.75$, $M_{\rm Pl} = 1.22 \times 10^{19}$ GeV. Coupling strength of the triplet fermion with the leptons of order $10^{-7}$ gives $K_{\Sigma_i} \sim 1$, which confirms the inverse decay not to come into thermal equilibrium. The Boltzmann equations for the evolution of triplet fermion and lepton number densities can be written in terms of yield parameter (i.e the ratio of number density to entropy density), which are provided by \cite{Plumacher:1996kc,Giudice:2003jh,Strumia:2006qk}
\begin{eqnarray}
&& s H z \frac{dY_{\Sigma}}{dz}=-\left(\frac{Y_\Sigma}{Y^{\rm eq}_\Sigma} - 1\right)\gamma_D -2 \left(\frac{Y^2_\Sigma}{(Y^{\rm eq}_\Sigma)^2} - 1\right)\gamma_A, \nn \\
&& s H z \frac{dY_L}{dz}= -\gamma_D \left(\frac{Y_\Sigma}{Y^{\rm eq}_\Sigma} - 1\right)\epsilon^{\Sigma}_{CP} -\frac{Y_L}{Y^{\rm eq}_L}\left(\frac{\gamma_D}{2} + 2\gamma_W \right).
\end{eqnarray}
where $s$ denotes the entropy density, $z = M_{\Sigma_i}/T$, $Y_{L} = Y_{\ell} -Y_{\overline{\ell}}$ and the equilibrium number densities are given by 
\begin{eqnarray}
Y^{\rm eq}_{\Sigma}= \frac{45  g_{\Sigma}}{4 {\pi}^4 g_\star} z^2 K_2(z), \hspace{3mm} {Y^{\rm eq}_\ell}= \frac{3}{4} \frac{45 \zeta(3) g_\ell}{2 {\pi}^4 g_{\star}}\,.
\end{eqnarray}
Here, $K_{1,2}$ denote the modified Bessel functions of type 1 and 2, $g_\ell=2$ and $g_{\Sigma}=2$ are the degrees of freedom of lepton and fermion triplets respectively. The reaction rate for the decay( $\gamma_D$) and gauge annihilation processes ($\gamma_A$) are given by  \cite{Hambye:2012fh}
\begin{equation}
\gamma_D = s Y^{\rm eq}_{\Sigma}\Gamma_{\Sigma} \frac{K_1(z)}{K_2(z)},~~ \gamma_A=\frac{M_{\Sigma_3} T^3}{32 \pi^3} e^{-2 z}\left[\frac{111 g^4}{8\pi}+\frac{3}{2 z}\left(\frac{111 g^4}{8\pi}+ \frac{51 g^4}{16 \pi} \right) + \mathcal{O}(1/z)^2 \right].
\end{equation}
The gauge annihilation processes of the decaying fermion includes $\Sigma \bar{\Sigma} \to f \bar{f}, G\bar{G}, H^\star H$, where $G$ stands for the gauge boson and $g$ is the usual gauge coupling. $\gamma_{w}$ are the lepton number violating washout processes ($\ell H \to {\bar{\ell}} {\bar{H}}$), which are suppressed due to the small coupling and can be safely neglected. We obtained a lepton asymmetry of order $\mathcal{O} (10^{-10})$ by solving the Boltzmann equations, represented in the left panel of Fig.\ref{Beqns} and the right panel clearly signifies the impact of decay, inverse decay and gauge scattering rates on the evolution of number densities with decrease in temperature.  
\section{Summary}
We explore the impact of modular $S_3$ symmetry on neutrino mixing and leptogenesis within the framework of type III seesaw. Since the usual $S_3$ symmetry requires number of scalar doublets, which leads to certain complications in explaining FCNCs and VEV alignments. We prefer the modular form of $S_3$, where the couplings transform non-trivially under the symmetry and replaces the requirement of multiple scalars. Since the scenario of type III seesaw is less frequented as compared to type I or type II, we explored a detailed analysis of neutrino mixing consistent with the $3\sigma$ observation. Numerical diagonalization of the flavored neutrino mass matrix provides an explanation of neutrino masses and mixing parameters in terms of the Yukawa couplings and free parameters. Thus we constrained all the model parameters from the neutrino oscillation data and the observed sum of active neutrino masses to obtain the correct ranges for the model predicted mixing angles and CP phase. We found the reactor mixing angle and Dirac CP phase to lie within the experimental limit. Apart, we also discussed the scenario of resonant leptogenesis by generating the lepton asymmetry from the decay of lightest fermion triplet to the final state lepton and Higgs. We solved the coupled Boltzmann equations to obtain the evolution of lepton asymmetric number density of required order ($Y_L \approx \mathcal{O} (10^{-10})$), which is adequate to generate an observed baryon asymmetry of order $Y_B \approx \mathcal{O}(10^{-11})$.   
\section*{Acknowledgments}
I acknowledge DST Inspire for its financial support. I thank to my supervisor Prof. Anjan Giri for his guidance and also thankful to Mitesh Behera and Dr. S. Singirala for their helpful discussions towards this work.


\bibliographystyle{utcaps_mod}
\bibliography{linear.bib}

\providecommand{\href}[2]{#2}\begingroup\raggedright\begin{thebibliography}{10}

\bibitem{Gripaios:2015gxa}
B.~Gripaios, ``{\em {Lectures on Physics Beyond the Standard Model}},''
  \href{http://arxiv.org/abs/1503.02636}{{\normalfont \ttfamily
  arXiv:1503.02636}}.

\bibitem{Araki:2004mb}
{\normalfont \bfseries KamLAND}, T.~Araki {\em et al.}, ``{\em {Measurement of
  neutrino oscillation with KamLAND: Evidence of spectral distortion}},''
  \href{http://dx.doi.org/10.1103/PhysRevLett.94.081801}{Phys. Rev. Lett.
  {\normalfont \bfseries 94} (2005)  081801},
  \href{http://arxiv.org/abs/hep-ex/0406035}{{\normalfont \ttfamily
  arXiv:hep-ex/0406035}}.

\bibitem{Tanabashi:2018oca}
{\normalfont \bfseries Particle Data Group}, M.~Tanabashi {\em et al.}, ``{\em
  {Review of Particle Physics}},''
\href{http://dx.doi.org/10.1103/PhysRevD.98.030001}{Phys. Rev. {\normalfont
  \bfseries D98} (2018) no.~3, 030001}.

\bibitem{Ade:2015xua}
{\normalfont \bfseries Planck}, P.~A.~R. Ade {\em et al.}, ``{\em {Planck 2015
  results. XIII. Cosmological parameters}},''
  \href{http://dx.doi.org/10.1051/0004-6361/201525830}{Astron. Astrophys.
  {\normalfont \bfseries 594} (2016)  A13},
\href{http://arxiv.org/abs/1502.01589}{{\normalfont \ttfamily
  arXiv:1502.01589}}.

\bibitem{King:2015bja}
S.~F. King, ``{\em {Discrete Symmetries and Models of Flavour Mixing}},''
\href{http://dx.doi.org/10.1088/1742-6596/631/1/012005}{J. Phys. Conf. Ser.
  {\normalfont \bfseries 631} (2015) no.~1, 012005}.

\bibitem{Chattopadhyay:2017zvs}
P.~Chattopadhyay and K.~M. Patel, ``{\em {Discrete symmetries for electroweak
  natural type-I seesaw mechanism}},''
  \href{http://dx.doi.org/10.1016/j.nuclphysb.2017.06.008}{Nucl. Phys.
  {\normalfont \bfseries B921} (2017)  487--506},
\href{http://arxiv.org/abs/1703.09541}{{\normalfont \ttfamily
  arXiv:1703.09541}}.

\bibitem{Sartori:1979gt}
G.~Sartori, ``{\em {Discrete Symmetries, Natural Flavor Conservation and Weak
  Mixing Angles}},''
\href{http://dx.doi.org/10.1016/0370-2693(79)90749-4}{Phys. Lett. {\normalfont
  \bfseries 82B} (1979)  255--259}.

\bibitem{Wilczek:1977uh}
F.~Wilczek and A.~Zee, ``{\em {Discrete Flavor Symmetries and a Formula for the
  Cabibbo Angle}},''
  \href{http://dx.doi.org/10.1016/0370-2693(77)90403-8}{Phys. Lett.
  {\normalfont \bfseries 70B} (1977)  418}.
[Erratum: Phys. Lett.72B,504(1978)].

\bibitem{DeRujula:1977dmn}
A.~De~Rujula, H.~Georgi, and S.~L. Glashow, ``{\em {A Theory of Flavor
  Mixing}},''
\href{http://dx.doi.org/10.1016/0003-4916(77)90172-5}{Annals Phys. {\normalfont
  \bfseries 109} (1977)  258}.

\bibitem{Haba:2005ds}
N.~Haba and K.~Yoshioka, ``{\em {Discrete flavor symmetry, dynamical mass
  textures, and grand unification}},''
  \href{http://dx.doi.org/10.1016/j.nuclphysb.2006.01.027}{Nucl. Phys.
  {\normalfont \bfseries B739} (2006)  254--284},
\href{http://arxiv.org/abs/hep-ph/0511108}{{\normalfont \ttfamily
  arXiv:hep-ph/0511108}}.

\bibitem{Altarelli:2010gt}
G.~Altarelli and F.~Feruglio, ``{\em {Discrete Flavor Symmetries and Models of
  Neutrino Mixing}},'' \href{http://dx.doi.org/10.1103/RevModPhys.82.2701}{Rev.
  Mod. Phys. {\normalfont \bfseries 82} (2010)  2701--2729},
  \href{http://arxiv.org/abs/1002.0211}{{\normalfont \ttfamily
  arXiv:1002.0211}}.

\bibitem{Petcov:2018snn}
S.~T. Petcov and A.~V. Titov, ``{\em {Assessing the Viability of $A_4$, $S_4$
  and $A_5$ Flavour Symmetries for Description of Neutrino Mixing}},''
  \href{http://dx.doi.org/10.1103/PhysRevD.97.115045}{Phys. Rev. {\normalfont
  \bfseries D97} (2018) no.~11, 115045},
\href{http://arxiv.org/abs/1804.00182}{{\normalfont \ttfamily
  arXiv:1804.00182}}.

\bibitem{Kubo:2004ps}
J.~Kubo, H.~Okada, and F.~Sakamaki, ``{\em {Higgs potential in minimal S(3)
  invariant extension of the standard model}},''
  \href{http://dx.doi.org/10.1103/PhysRevD.70.036007}{Phys. Rev. {\normalfont
  \bfseries D70} (2004)  036007},
\href{http://arxiv.org/abs/hep-ph/0402089}{{\normalfont \ttfamily
  arXiv:hep-ph/0402089}}.

\bibitem{Borah:2017dmk}
D.~Borah and B.~Karmakar, ``{\em {$A_4$ flavour model for Dirac neutrinos: Type
  I and inverse seesaw}},''
  \href{http://dx.doi.org/10.1016/j.physletb.2018.03.047}{Phys. Lett. B
  {\normalfont \bfseries 780} (2018)  461--470},
  \href{http://arxiv.org/abs/1712.06407}{{\normalfont \ttfamily
  arXiv:1712.06407}}.

\bibitem{Ishimori:2010au}
H.~Ishimori, T.~Kobayashi, H.~Ohki, Y.~Shimizu, H.~Okada, and M.~Tanimoto,
  ``{\em {Non-Abelian Discrete Symmetries in Particle Physics}},''
  \href{http://dx.doi.org/10.1143/PTPS.183.1}{Prog. Theor. Phys. Suppl.
  {\normalfont \bfseries 183} (2010)  1--163},
  \href{http://arxiv.org/abs/1003.3552}{{\normalfont \ttfamily
  arXiv:1003.3552}}.

\bibitem{Feruglio:2017spp}
F.~Feruglio, {\em {Are neutrino masses modular forms?}},
  \href{http://dx.doi.org/10.1142/9789813238053\_0012}{pp.~227--266}.
\newblock 2019.
\newblock \href{http://arxiv.org/abs/1706.08749}{{\normalfont \ttfamily
  arXiv:1706.08749}}.

\bibitem{Acharya:1995ag}
B.~S. Acharya, D.~Bailin, A.~Love, W.~Sabra, and S.~Thomas, ``{\em {Spontaneous
  breaking of CP symmetry by orbifold moduli}},''
  \href{http://dx.doi.org/10.1016/0370-2693(95)00945-H}{Phys. Lett. B
  {\normalfont \bfseries 357} (1995)  387--396},
  \href{http://arxiv.org/abs/hep-th/9506143}{{\normalfont \ttfamily
  arXiv:hep-th/9506143}}. [Erratum: Phys.Lett.B 407, 451--451 (1997)].

\bibitem{Lu:2019vgm}
J.-N. Lu, X.-G. Liu, and G.-J. Ding, ``{\em {Modular symmetry origin of texture
  zeros and quark lepton unification}},''
  \href{http://dx.doi.org/10.1103/PhysRevD.101.115020}{Phys. Rev. D
  {\normalfont \bfseries 101} (2020) no.~11, 115020},
  \href{http://arxiv.org/abs/1912.07573}{{\normalfont \ttfamily
  arXiv:1912.07573}}.

\bibitem{Novichkov:2019sqv}
P.~Novichkov, J.~Penedo, S.~Petcov, and A.~Titov, ``{\em {Generalised CP
  Symmetry in Modular-Invariant Models of Flavour}},''
  \href{http://dx.doi.org/10.1007/JHEP07(2019)165}{JHEP {\normalfont \bfseries
  07} (2019)  165}, \href{http://arxiv.org/abs/1905.11970}{{\normalfont
  \ttfamily arXiv:1905.11970}}.

\bibitem{Baur:2019kwi}
A.~Baur, H.~P. Nilles, A.~Trautner, and P.~K. Vaudrevange, ``{\em {Unification
  of Flavor, CP, and Modular Symmetries}},''
  \href{http://dx.doi.org/10.1016/j.physletb.2019.03.066}{Phys. Lett. B
  {\normalfont \bfseries 795} (2019)  7--14},
  \href{http://arxiv.org/abs/1901.03251}{{\normalfont \ttfamily
  arXiv:1901.03251}}.

\bibitem{Dent:2001cc}
T.~Dent, ``{\em {CP violation and modular symmetries}},''
  \href{http://dx.doi.org/10.1103/PhysRevD.64.056005}{Phys. Rev. D {\normalfont
  \bfseries 64} (2001)  056005},
  \href{http://arxiv.org/abs/hep-ph/0105285}{{\normalfont \ttfamily
  arXiv:hep-ph/0105285}}.

\bibitem{Giedt:2002ns}
J.~Giedt, ``{\em {CP violation and moduli stabilization in heterotic
  models}},'' \href{http://dx.doi.org/10.1142/S0217732302007879}{Mod. Phys.
  Lett. A {\normalfont \bfseries 17} (2002)  1465--1473},
  \href{http://arxiv.org/abs/hep-ph/0204017}{{\normalfont \ttfamily
  arXiv:hep-ph/0204017}}.

\bibitem{Chen:2019ewa}
M.-C. Chen, S.~Ramos-Sánchez, and M.~Ratz, ``{\em {A note on the predictions
  of models with modular flavor symmetries}},''
  \href{http://dx.doi.org/10.1016/j.physletb.2019.135153}{Phys. Lett. B
  {\normalfont \bfseries 801} (2020)  135153},
  \href{http://arxiv.org/abs/1909.06910}{{\normalfont \ttfamily
  arXiv:1909.06910}}.

\bibitem{King:2020qaj}
S.~J. King and S.~F. King, ``{\em {Fermion Mass Hierarchies from Modular
  Symmetry}},'' \href{http://arxiv.org/abs/2002.00969}{{\normalfont \ttfamily
  arXiv:2002.00969}}.

\bibitem{Nomura:2019xsb}
T.~Nomura, H.~Okada, and S.~Patra, ``{\em {An Inverse Seesaw model with
  $A_4$-modular symmetry}},''
  \href{http://arxiv.org/abs/1912.00379}{{\normalfont \ttfamily
  arXiv:1912.00379}}.

\bibitem{Feruglio:2017ieh}
F.~Feruglio, ``{\em {Neutrino masses and mixing angles: A tribute to Guido
  Altarelli}},'' Frascati Phys. Ser. {\normalfont \bfseries 64} (2017)
  174--182.

\bibitem{Behera:2020sfe}
M.~K. Behera, S.~Mishra, S.~Singirala, and R.~Mohanta, ``{\em {Implications of
  $A_4$ modular symmetry on Neutrino mass, Mixing and Leptogenesis with Linear
  Seesaw}},'' \href{http://arxiv.org/abs/2007.00545}{{\normalfont \ttfamily
  arXiv:2007.00545}}.

\bibitem{Kobayashi:2019xvz}
T.~Kobayashi, Y.~Shimizu, K.~Takagi, M.~Tanimoto, and T.~H. Tatsuishi, ``{\em
  {$A_4$ lepton flavor model and modulus stabilization from $S_4$ modular
  symmetry}},'' \href{http://dx.doi.org/10.1103/PhysRevD.100.115045}{Phys. Rev.
  D {\normalfont \bfseries 100} (2019) no.~11, 115045},
  \href{http://arxiv.org/abs/1909.05139}{{\normalfont \ttfamily
  arXiv:1909.05139}}. [Erratum: Phys.Rev.D 101, 039904 (2020)].

\bibitem{Abbas:2020qzc}
M.~Abbas, ``{\em {Flavor masses and mixing in modular $A$$_{4}$ Symmetry}},''
  \href{http://arxiv.org/abs/2002.01929}{{\normalfont \ttfamily
  arXiv:2002.01929}}.

\bibitem{Wang:2019xbo}
X.~Wang, ``{\em {Lepton Flavor Mixing and CP Violation in the Minimal
  Type-(I+II) Seesaw Model with a Modular $A_4$ Symmetry}},''
  \href{http://arxiv.org/abs/1912.13284}{{\normalfont \ttfamily
  arXiv:1912.13284}}.

\bibitem{Kobayashi:2019gtp}
T.~Kobayashi, T.~Nomura, and T.~Shimomura, ``{\em {Type II seesaw models with
  modular $A_4$ symmetry}},''
  \href{http://arxiv.org/abs/1912.00637}{{\normalfont \ttfamily
  arXiv:1912.00637}}.

\bibitem{Penedo:2018nmg}
J.~Penedo and S.~Petcov, ``{\em {Lepton Masses and Mixing from Modular $S_4$
  Symmetry}},'' \href{http://dx.doi.org/10.1016/j.nuclphysb.2018.12.016}{Nucl.\
  Phys.\ B {\normalfont \bfseries 939} (2019)  292--307},
  \href{http://arxiv.org/abs/1806.11040}{{\normalfont \ttfamily
  arXiv:1806.11040}}.

\bibitem{Liu:2020akv}
X.-G. Liu, C.-Y. Yao, and G.-J. Ding, ``{\em {Modular Invariant Quark and
  Lepton Models in Double Covering of $S_4$ Modular Group}},''
  \href{http://arxiv.org/abs/2006.10722}{{\normalfont \ttfamily
  arXiv:2006.10722}}.

\bibitem{Gui-JunDing:2019wap}
G.-J. Ding, S.~F. King, X.-G. Liu, and J.-N. Lu, ``{\em {Modular S$_{4}$ and
  A$_{4}$ symmetries and their fixed points: new predictive examples of lepton
  mixing}},'' \href{http://dx.doi.org/10.1007/JHEP12(2019)030}{JHEP
  {\normalfont \bfseries 12} (2019)  030},
  \href{http://arxiv.org/abs/1910.03460}{{\normalfont \ttfamily
  arXiv:1910.03460}}.

\bibitem{Ding:2019zxk}
G.-J. Ding, S.~F. King, and X.-G. Liu, ``{\em {Modular A$_{4}$ symmetry models
  of neutrinos and charged leptons}},''
  \href{http://dx.doi.org/10.1007/JHEP09(2019)074}{JHEP {\normalfont \bfseries
  09} (2019)  074}, \href{http://arxiv.org/abs/1907.11714}{{\normalfont
  \ttfamily arXiv:1907.11714}}.

\bibitem{Novichkov:2018nkm}
P.~Novichkov, J.~Penedo, S.~Petcov, and A.~Titov, ``{\em {Modular A$_{5}$
  symmetry for flavour model building}},''
  \href{http://dx.doi.org/10.1007/JHEP04(2019)174}{JHEP {\normalfont \bfseries
  04} (2019)  174}, \href{http://arxiv.org/abs/1812.02158}{{\normalfont
  \ttfamily arXiv:1812.02158}}.

\bibitem{Kubo:2003pd}
J.~Kubo, ``{\em {Majorana phase in minimal S(3) invariant extension of the
  standard model}},'' \href{http://dx.doi.org/10.1016/j.physletb.2005.06.013,
  10.1016/j.physletb.2003.10.048}{Phys. Lett. {\normalfont \bfseries B578}
  (2004)  156--164}, \href{http://arxiv.org/abs/hep-ph/0309167}{{\normalfont
  \ttfamily arXiv:hep-ph/0309167}}.
[Erratum: Phys. Lett.B619,387(2005)].

\bibitem{Meloni:2010aw}
D.~Meloni, S.~Morisi, and E.~Peinado, ``{\em {Fritzsch neutrino mass matrix
  from $S_3$ symmetry}},''
  \href{http://dx.doi.org/10.1088/0954-3899/38/1/015003}{J. Phys. {\normalfont
  \bfseries G38} (2011)  015003},
\href{http://arxiv.org/abs/1005.3482}{{\normalfont \ttfamily arXiv:1005.3482}}.

\bibitem{Canales:2012dr}
F.~Gonzalez~Canales, A.~Mondragon, and M.~Mondragon, ``{\em {The $S_3$ Flavour
  Symmetry: Neutrino Masses and Mixings}},''
  \href{http://dx.doi.org/10.1002/prop.201200121}{Fortsch. Phys. {\normalfont
  \bfseries 61} (2013)  546--570},
\href{http://arxiv.org/abs/1205.4755}{{\normalfont \ttfamily arXiv:1205.4755}}.

\bibitem{Mondragon:2007af}
A.~Mondragon, M.~Mondragon, and E.~Peinado, ``{\em {Lepton masses, mixings and
  FCNC in a minimal S(3)-invariant extension of the Standard Model}},''
  \href{http://dx.doi.org/10.1103/PhysRevD.76.076003}{Phys. Rev. {\normalfont
  \bfseries D76} (2007)  076003},
\href{http://arxiv.org/abs/0706.0354}{{\normalfont \ttfamily arXiv:0706.0354}}.

\bibitem{Araki:2005ec}
T.~Araki, J.~Kubo, and E.~A. Paschos, ``{\em {S(3) flavor symmetry and
  leptogenesis}},'' \href{http://dx.doi.org/10.1140/epjc/s2005-02434-3}{Eur.
  Phys. J. {\normalfont \bfseries C45} (2006)  465--475},
\href{http://arxiv.org/abs/hep-ph/0502164}{{\normalfont \ttfamily
  arXiv:hep-ph/0502164}}.

\bibitem{Kobayashi:2019rzp}
T.~Kobayashi, Y.~Shimizu, K.~Takagi, M.~Tanimoto, and T.~H. Tatsuishi, ``{\em
  {Modular $S_3$ invariant flavor model in SU(5) GUT}},''
  \href{http://arxiv.org/abs/1906.10341}{{\normalfont \ttfamily
  arXiv:1906.10341}}.

\bibitem{Davidson:2008bu}
S.~Davidson, E.~Nardi, and Y.~Nir, ``{\em {Leptogenesis}},''
  \href{http://dx.doi.org/10.1016/j.physrep.2008.06.002}{Phys. Rept.
  {\normalfont \bfseries 466} (2008)  105--177},
  \href{http://arxiv.org/abs/0802.2962}{{\normalfont \ttfamily
  arXiv:0802.2962}}.

\bibitem{Weinberg:1980bf}
S.~Weinberg, ``{\em {Varieties of Baryon and Lepton Nonconservation}},''
\href{http://dx.doi.org/10.1103/PhysRevD.22.1694}{Phys. Rev. {\normalfont
  \bfseries D22} (1980)  1694}.

\bibitem{Pascoli:2006ie}
S.~Pascoli, S.~T. Petcov, and A.~Riotto, ``{\em {Connecting low energy leptonic
  CP-violation to leptogenesis}},''
  \href{http://dx.doi.org/10.1103/PhysRevD.75.083511}{Phys. Rev. {\normalfont
  \bfseries D75} (2007)  083511},
\href{http://arxiv.org/abs/hep-ph/0609125}{{\normalfont \ttfamily
  arXiv:hep-ph/0609125}}.

\bibitem{Sierra:2014tqa}
D.~Aristizabal~Sierra, M.~Dhen, and T.~Hambye, ``{\em {Scalar triplet flavored
  leptogenesis: a systematic approach}},''
  \href{http://dx.doi.org/10.1088/1475-7516/2014/08/003}{JCAP {\normalfont
  \bfseries 1408} (2014)  003},
\href{http://arxiv.org/abs/1401.4347}{{\normalfont \ttfamily arXiv:1401.4347}}.

\bibitem{Felipe:2013kk}
R.~Gonzalez~Felipe, F.~R. Joaquim, and H.~Serodio, ``{\em {Flavoured CP
  asymmetries for type II seesaw leptogenesis}},''
  \href{http://dx.doi.org/10.1142/S0217751X13501650}{Int. J. Mod. Phys.
  {\normalfont \bfseries A28} (2013)  1350165},
\href{http://arxiv.org/abs/1301.0288}{{\normalfont \ttfamily arXiv:1301.0288}}.

\bibitem{Hambye:2003ka}
T.~Hambye and G.~Senjanovic, ``{\em {Consequences of triplet seesaw for
  leptogenesis}},''
  \href{http://dx.doi.org/10.1016/j.physletb.2003.11.061}{Phys. Lett.
  {\normalfont \bfseries B582} (2004)  73--81},
\href{http://arxiv.org/abs/hep-ph/0307237}{{\normalfont \ttfamily
  arXiv:hep-ph/0307237}}.

\bibitem{Senami:2003jn}
M.~Senami and K.~Yamamoto, ``{\em {Leptogenesis with supersymmetric Higgs
  triplets in TeV region}},''
  \href{http://dx.doi.org/10.1142/S0217751X06029478}{Int. J. Mod. Phys.
  {\normalfont \bfseries A21} (2006)  1291--1306},
\href{http://arxiv.org/abs/hep-ph/0305202}{{\normalfont \ttfamily
  arXiv:hep-ph/0305202}}.

\bibitem{Lavignac:2015gpa}
S.~Lavignac and B.~Schmauch, ``{\em {Flavour always matters in scalar triplet
  leptogenesis}},'' \href{http://dx.doi.org/10.1007/JHEP05(2015)124}{JHEP
  {\normalfont \bfseries 05} (2015)  124},
\href{http://arxiv.org/abs/1503.00629}{{\normalfont \ttfamily
  arXiv:1503.00629}}.

\bibitem{AristizabalSierra:2012pv}
D.~Aristizabal~Sierra, ``{\em {Scalar triplet leptogenesis without right-handed
  neutrino decoupling}},''
  \href{http://dx.doi.org/10.1016/j.nuclphysbps.2013.04.054}{Nucl. Phys. Proc.
  Suppl. {\normalfont \bfseries 237-238} (2013)  43--45},
\href{http://arxiv.org/abs/1212.3302}{{\normalfont \ttfamily arXiv:1212.3302}}.

\bibitem{Chen:2009vx}
S.-L. Chen and X.-G. He, ``{\em {Leptogenesis and LHC Physics with Type III
  See-Saw}},'' \href{http://dx.doi.org/10.1142/S2010194511000067}{Int. J. Mod.
  Phys. Conf. Ser. {\normalfont \bfseries 01} (2011)  18--27},
  \href{http://arxiv.org/abs/0901.1264}{{\normalfont \ttfamily
  arXiv:0901.1264}}.

\bibitem{Kobayashi:2018wkl}
T.~Kobayashi, Y.~Shimizu, K.~Takagi, M.~Tanimoto, T.~H. Tatsuishi, and
  H.~Uchida, ``{\em {Finite modular subgroups for fermion mass matrices and
  baryon/lepton number violation}},''
  \href{http://dx.doi.org/10.1016/j.physletb.2019.05.034}{Phys. Lett. B
  {\normalfont \bfseries 794} (2019)  114--121},
  \href{http://arxiv.org/abs/1812.11072}{{\normalfont \ttfamily
  arXiv:1812.11072}}.

\bibitem{deAdelhartToorop:2011re}
R.~de~Adelhart~Toorop, F.~Feruglio, and C.~Hagedorn, ``{\em {Finite Modular
  Groups and Lepton Mixing}},''
  \href{http://dx.doi.org/10.1016/j.nuclphysb.2012.01.017}{Nucl.\ Phys.\ B
  {\normalfont \bfseries 858} (2012)  437--467},
  \href{http://arxiv.org/abs/1112.1340}{{\normalfont \ttfamily
  arXiv:1112.1340}}.

\bibitem{Okada:2019xqk}
H.~Okada and Y.~Orikasa, ``{\em {Modular $S_3$ symmetric radiative seesaw
  model}},'' \href{http://dx.doi.org/10.1103/PhysRevD.100.115037}{Phys. Rev. D
  {\normalfont \bfseries 100} (2019) no.~11, 115037},
  \href{http://arxiv.org/abs/1907.04716}{{\normalfont \ttfamily
  arXiv:1907.04716}}.

\bibitem{Hambye:2013jsa}
T.~Hambye, ``{\em {CLFV and the origin of neutrino masses}},''
  \href{http://dx.doi.org/10.1016/j.nuclphysbps.2014.02.004}{Nucl. Phys. B
  Proc. Suppl. {\normalfont \bfseries 248-250} (2014)  13--19},
  \href{http://arxiv.org/abs/1312.5214}{{\normalfont \ttfamily
  arXiv:1312.5214}}.

\bibitem{Bandyopadhyay:2009xa}
P.~Bandyopadhyay, S.~Choubey, and M.~Mitra, ``{\em {Two Higgs Doublet Type III
  Seesaw with mu-tau symmetry at LHC}},''
  \href{http://dx.doi.org/10.1088/1126-6708/2009/10/012}{JHEP {\normalfont
  \bfseries 10} (2009)  012},
  \href{http://arxiv.org/abs/0906.5330}{{\normalfont \ttfamily
  arXiv:0906.5330}}.

\bibitem{Esteban:2018azc}
I.~Esteban, M.~Gonzalez-Garcia, A.~Hernandez-Cabezudo, M.~Maltoni, and
  T.~Schwetz, ``{\em {Global analysis of three-flavour neutrino oscillations:
  synergies and tensions in the determination of $\theta_{23}$, $\delta_{CP}$,
  and the mass ordering}},''
  \href{http://dx.doi.org/10.1007/JHEP01(2019)106}{JHEP {\normalfont \bfseries
  01} (2019)  106}, \href{http://arxiv.org/abs/1811.05487}{{\normalfont
  \ttfamily arXiv:1811.05487}}.

\bibitem{Aghanim:2018eyx}
{\normalfont \bfseries Planck}, N.~Aghanim {\em et al.}, ``{\em {Planck 2018
  results. VI. Cosmological parameters}},''
  \href{http://arxiv.org/abs/1807.06209}{{\normalfont \ttfamily
  arXiv:1807.06209}}.

\bibitem{Buchmuller:2004nz}
W.~Buchmuller, P.~Di~Bari, and M.~Plumacher, ``{\em {Leptogenesis for
  pedestrians}},'' \href{http://dx.doi.org/10.1016/j.aop.2004.02.003}{Annals
  Phys. {\normalfont \bfseries 315} (2005)  305--351},
  \href{http://arxiv.org/abs/hep-ph/0401240}{{\normalfont \ttfamily
  arXiv:hep-ph/0401240}}.

\bibitem{Asaka:2018hyk}
T.~Asaka and T.~Yoshida, ``{\em {Resonant leptogenesis at TeV-scale and
  neutrinoless double beta decay}},''
  \href{http://dx.doi.org/10.1007/JHEP09(2019)089}{JHEP {\normalfont \bfseries
  09} (2019)  089}, \href{http://arxiv.org/abs/1812.11323}{{\normalfont
  \ttfamily arXiv:1812.11323}}.

\bibitem{Abada:2018oly}
A.~Abada, G.~Arcadi, V.~Domcke, M.~Drewes, J.~Klaric, and M.~Lucente, ``{\em
  {Low-scale leptogenesis with three heavy neutrinos}},''
  \href{http://dx.doi.org/10.1007/JHEP01(2019)164}{JHEP {\normalfont \bfseries
  01} (2019)  164}, \href{http://arxiv.org/abs/1810.12463}{{\normalfont
  \ttfamily arXiv:1810.12463}}.

\bibitem{Pilaftsis:2003gt}
A.~Pilaftsis and T.~E. Underwood, ``{\em {Resonant leptogenesis}},''
  \href{http://dx.doi.org/10.1016/j.nuclphysb.2004.05.029}{Nucl. Phys. B
  {\normalfont \bfseries 692} (2004)  303--345},
  \href{http://arxiv.org/abs/hep-ph/0309342}{{\normalfont \ttfamily
  arXiv:hep-ph/0309342}}.

\bibitem{Iso:2010mv}
S.~Iso, N.~Okada, and Y.~Orikasa, ``{\em {Resonant Leptogenesis in the Minimal
  B-L Extended Standard Model at TeV}},''
  \href{http://dx.doi.org/10.1103/PhysRevD.83.093011}{Phys. Rev. D {\normalfont
  \bfseries 83} (2011)  093011},
  \href{http://arxiv.org/abs/1011.4769}{{\normalfont \ttfamily
  arXiv:1011.4769}}.

\bibitem{Adhikary:2014qba}
B.~Adhikary, M.~Chakraborty, and A.~Ghosal, ``{\em {Flavored leptogenesis with
  quasidegenerate neutrinos in a broken cyclic symmetric model}},''
  \href{http://dx.doi.org/10.1103/PhysRevD.93.113001}{Phys. Rev. {\normalfont
  \bfseries D93} (2016) no.~11, 113001},
\href{http://arxiv.org/abs/1407.6173}{{\normalfont \ttfamily arXiv:1407.6173}}.

\bibitem{Pilaftsis:2005rv}
A.~Pilaftsis and T.~E. Underwood, ``{\em {Electroweak-scale resonant
  leptogenesis}},'' \href{http://dx.doi.org/10.1103/PhysRevD.72.113001}{Phys.
  Rev. D {\normalfont \bfseries 72} (2005)  113001},
  \href{http://arxiv.org/abs/hep-ph/0506107}{{\normalfont \ttfamily
  arXiv:hep-ph/0506107}}.

\bibitem{Albright:2003xb}
C.~H. Albright and S.~Barr, ``{\em {Leptogenesis in the type III seesaw
  mechanism}},'' \href{http://dx.doi.org/10.1103/PhysRevD.69.073010}{Phys. Rev.
  D {\normalfont \bfseries 69} (2004)  073010},
  \href{http://arxiv.org/abs/hep-ph/0312224}{{\normalfont \ttfamily
  arXiv:hep-ph/0312224}}.

\bibitem{Goswami:2018jar}
S.~Goswami, K.~Vishnudath, and N.~Khan, ``{\em {Constraining the minimal
  type-III seesaw model with naturalness, lepton flavor violation, and
  electroweak vacuum stability}},''
  \href{http://dx.doi.org/10.1103/PhysRevD.99.075012}{Phys. Rev. D {\normalfont
  \bfseries 99} (2019) no.~7, 075012},
  \href{http://arxiv.org/abs/1810.11687}{{\normalfont \ttfamily
  arXiv:1810.11687}}.

\bibitem{Franceschini:2008pz}
R.~Franceschini, T.~Hambye, and A.~Strumia, ``{\em {Type-III see-saw at
  LHC}},'' \href{http://dx.doi.org/10.1103/PhysRevD.78.033002}{Phys. Rev. D
  {\normalfont \bfseries 78} (2008)  033002},
  \href{http://arxiv.org/abs/0805.1613}{{\normalfont \ttfamily
  arXiv:0805.1613}}.

\bibitem{Pilaftsis:1997jf}
A.~Pilaftsis, ``{\em {CP violation and baryogenesis due to heavy Majorana
  neutrinos}},'' \href{http://dx.doi.org/10.1103/PhysRevD.56.5431}{Phys. Rev. D
  {\normalfont \bfseries 56} (1997)  5431--5451},
  \href{http://arxiv.org/abs/hep-ph/9707235}{{\normalfont \ttfamily
  arXiv:hep-ph/9707235}}.

\bibitem{Hambye:2012fh}
T.~Hambye, ``{\em {Leptogenesis: beyond the minimal type I seesaw scenario}},''
  \href{http://dx.doi.org/10.1088/1367-2630/14/12/125014}{New J. Phys.
  {\normalfont \bfseries 14} (2012)  125014},
\href{http://arxiv.org/abs/1212.2888}{{\normalfont \ttfamily arXiv:1212.2888}}.

\bibitem{Sakharov:1967dj}
A.~Sakharov, ``{\em {Violation of CP Invariance, C asymmetry, and baryon
  asymmetry of the universe}},''
  \href{http://dx.doi.org/10.1070/PU1991v034n05ABEH002497}{Sov. Phys. Usp.
  {\normalfont \bfseries 34} (1991) no.~5, 392--393}.

\bibitem{Plumacher:1996kc}
M.~Plumacher, ``{\em {Baryogenesis and lepton number violation}},''
  \href{http://dx.doi.org/10.1007/s002880050418}{Z. Phys. C {\normalfont
  \bfseries 74} (1997)  549--559},
  \href{http://arxiv.org/abs/hep-ph/9604229}{{\normalfont \ttfamily
  arXiv:hep-ph/9604229}}.

\bibitem{Giudice:2003jh}
G.~Giudice, A.~Notari, M.~Raidal, A.~Riotto, and A.~Strumia, ``{\em {Towards a
  complete theory of thermal leptogenesis in the SM and MSSM}},''
  \href{http://dx.doi.org/10.1016/j.nuclphysb.2004.02.019}{Nucl. Phys. B
  {\normalfont \bfseries 685} (2004)  89--149},
  \href{http://arxiv.org/abs/hep-ph/0310123}{{\normalfont \ttfamily
  arXiv:hep-ph/0310123}}.

\bibitem{Strumia:2006qk}
A.~Strumia, ``{\em {Baryogenesis via leptogenesis}},'' in {\em {Les Houches
  Summer School on Theoretical Physics: Session 84: Particle Physics Beyond the
  Standard Model}}, pp.~655--680.
\newblock 8, 2006.
\newblock \href{http://arxiv.org/abs/hep-ph/0608347}{{\normalfont \ttfamily
  arXiv:hep-ph/0608347}}.

\end{thebibliography}\endgroup
\end{document}